\documentclass[fleqn,usenatbib]{mnras}

\usepackage{newtxtext,newtxmath}
\usepackage[T1]{fontenc}

\DeclareRobustCommand{\VAN}[3]{#2}
\let\VANthebibliography\thebibliography
\def\thebibliography{\DeclareRobustCommand{\VAN}[3]{##3}\VANthebibliography}

\usepackage{appendix}
\usepackage{graphicx}
\usepackage{dcolumn}
\usepackage{bm}
\usepackage{amsmath}
\usepackage{float}
\usepackage{lipsum}
\usepackage{color}
\usepackage{times}
\usepackage{textcomp}
\usepackage{latexsym}
\usepackage{hyperref}

\newcommand{\HU}{\mathcal{H}_U}
\newcommand{\HL}{\mathcal{H}_L}

\newcommand{\HLT}{\mathcal{H}_L^{\rm Tidal}}
\newcommand{\HLDNS}{\mathcal{H}_L^{\rm DNS}}	% Extra maths symbols

\title[Binary Neutron Stars Lensing]{Lensed or not lensed: Determining lensing magnifications for binary neutron star mergers from a single detection}

\author[P. T. H. Pang et al.]{
Peter T. H. Pang$^{1,2}$\thanks{thopang@nikhef.nl},
Otto A. Hannuksela$^{1,2}$\thanks{o.hannuksela@nikhef.nl},
Tim Dietrich$^{3,1}$,
Giulia Pagano$^{4,5}$,
Ian W. Harry$^{6}$
\\
$^{1}$Nikhef – National Institute for Subatomic Physics, Science Park, 1098 XG Amsterdam, The Netherlands\\
$^{2}$Department of Physics, Utrecht University, Princetonplein 1, 3584 CC Utrecht, The Netherlands\\
$^{3}$Institute for Physics and Astronomy, University of Potsdam,
Karl-Liebknecht-Str. 24/25, 14476, Potsdam, Germany\\
$^{4}$Dipartimento di Fisica “Enrico Fermi”, Università di Pisa, Pisa I-56127, Italy\\
$^{5}$INFN sezione di Pisa, Pisa I-56127, Italy\\
$^{6}$Institute for Cosmology and Gravitation, University of Portsmouth, 1-8 Burnaby Road, Portsmouth, P01 3FZ, UK
}

\pubyear{2020}

\begin{document}
\label{firstpage}
\pagerange{\pageref{firstpage}--\pageref{lastpage}}
\maketitle

\begin{abstract}
\noindent
Advanced LIGO and Advanced Virgo could observe the first lensed gravitational wave sources in the coming years, while the future Einstein Telescope could observe hundreds of lensed events. It is, therefore, crucial to develop methodologies to distinguish between lensed from unlensed gravitational-wave observations. A lensed signal not identified as such will lead to biases during the interpretation of the source. In particular, sources will appear to have intrinsically higher masses. No robust method currently exists to distinguish between the magnification bias caused by lensing and intrinsically high-mass sources. In this work, we show how to recognize lensed and unlensed binary neutron star systems through the measurement of their tidal effects for highly magnified sources as a proof-of-principle. The proposed method could be used to identify lensed binary neutron stars, which are the chief candidate for lensing cosmography studies. We apply our method on GW190425, finding no evidence in favor of lensing, mainly due to the poor measurement of the event's tidal effects. However, we expect that future detections with better tidal measurements can yield better constraints.
\end{abstract}

\begin{keywords}
Gravitational lensing: strong -- Neutron star merger -- Gravitational waves
\end{keywords}

%%%%%%%%%%%%%%%%%%%%%%%%%%%%%%%%%%%%%%%%%%%%%%%%%%

%%%%%%%%%%%%%%%%% BODY OF PAPER %%%%%%%%%%%%%%%%%%

\section{Introduction} 

Between 2015 and 2017, Advanced LIGO~\cite{AdvLIGO2015} and Advanced Virgo~\cite{Acernese_2014} conducted their first two observation runs (O1 and O2) detecting several binary
black hole (BBH) mergers and one binary neutron star (BNS)
merger~\cite{PhysRevX.9.031040}. 
The third observation run (O3) is currently on-going and numerous candidate
gravitational transients have been observed~\cite{GCN,graceDB}.
Future observing runs will see upgrades to the Advanced LIGO detectors and the Advanced Virgo detector, and, in addition, the Japanese observatory KAGRA~\cite{Somiya_2012,PhysRevD.88.043007,10.1093/ptep/ptx180} is expected to join the network in 2020~\cite{collaboration2013prospects}.

When gravitational waves (GWs) travel near a galaxy or a galaxy cluster, their trajectories are curved, resulting in strong gravitational lensing~\cite{ohanian1974focusing,bliokh1975diffraction,bontz1981diffraction,thorne1983theory,deguchi1986diffraction,nakamura1998gravitational,Takahashi_2003,2019RPPh...82l6901O}. 
The lensing magnifies the amplitude of the waves without changing their frequency evolution~\cite{PhysRevLett.77.2875,PhysRevD.95.044011}.
In the case of strong lensing by galaxies, it is possible to produce multiple ``images'', which would arrive to us with relative time-delays between minutes and weeks\footnote{Let us note gravitational lensing by galaxy clusters could produce time-delays as large as months to years~\cite{Smith:2017mqu,smith2018strong}. However, we do not consider this scenario here. }~\cite{2011MNRAS.415.2773S,Haris:2018vmn}. 
Based on predictions on the number of expected GW sources, and the distribution of lenses in the Universe, Refs.~\cite{PhysRevD.97.023012,Li_2018,2018MNRAS.480.3842O} suggest that lensed gravitational-waves may be detected in the coming years, as the LIGO/Virgo detectors reach their design sensitivities.\footnote{Specifically, Refs.~\cite{PhysRevD.97.023012,Li_2018,2018MNRAS.480.3842O} arrive at $\sim 0.1 - 10 \, \rm yr^{-1}$ observable lensed events per year.}
The number of detectable events could reach hundreds in the Einstein Telescope~\cite{2014JCAP...10..080B,2015JCAP...12..006D}.
Lensed GWs present several potential applications in fundamental physics, astrophysics, and cosmology~\cite{2011MNRAS.415.2773S,PhysRevD.98.083005,PhysRevD.86.022004,PhysRevLett.118.091101,PhysRevD.95.063512,PhysRevLett.118.091102,2018EPJC...78..692Y,2019arXiv190808950M}.

A number of possibilities to identify a lensed GW signal have been proposed. 
One can look for signatures of multiple images or microlensing within GW data~\cite{Haris:2018vmn, PhysRevD.98.083005, PhysRevD.98.103022, 2018PhRvD..98j4029D, 2019arXiv190406020L, 2019arXiv191205389M,PhysRevD.101.024039}. 
Alternatively, one could search for a population of apparently high-mass binary events produced by lensing magnification~\cite{PhysRevD.95.044011,broadhurst2018reinterpreting,2018MNRAS.480.3842O}. 
The first combined search for all these signatures was performed recently on the O1/O2 data~\cite{Hannuksela_2019}.

Here we focus on the problem of reliably identifying lensed binary
neutron star signals.
The overall magnification caused by lensing is degenerate with the
luminosity distance measured from the GW signal and so a lensed system will
appear to be closer than it truly is~\cite{PhysRevD.95.044011,PhysRevD.97.023012,broadhurst2018reinterpreting,2018MNRAS.480.3842O,Contigiani:2020yyc}.
As the distance to the binary is biased, the estimation of the redshift
to the binary will be as well. A redshifted gravitational-wave signal will
appear to an observer to have higher masses than in reality.

The recent high-mass BNS detection,
GW190425~\cite{collaboration2020gw190425}, is therefore of particular
interest
The mass of the system is higher than expected from the known galactic double neutron star population~\cite{Farrow_2019,2020arXiv200104502S}.
Could this signal be a lensed system consistent with the known
population?
Unfortunately, to answer this question definitively, we would need a unique signature
to discern an intrinsically high-mass binary event from a lensed event.

We note that the problem could, in principle, be resolved by lens statistics: 
the lensed hypothesis is disfavored \textit{a priori}, as the rate of lensed BNSs is low within LIGO/Virgo~\cite{2018MNRAS.480.3842O,smith2019discovery}. 
However, the prior probability of the other hypothesis (a new population of BNSs) is largely unknown, as the masses are inconsistent with the observed double neutron star population within the galaxy~\cite{Farrow_2019,2020arXiv200104502S}. 
Without a good grasp of the relative prior probabilities of the two hypotheses, a quantitative comparison is challenging. 

In this work, we propose a robust method to rule out or confirm the lensing hypothesis for BNSs.
While GW lensing biases the intrinsic mass measurement, it does not bias the tidal deformabilities as measured from the GW phasing. 
Therefore, a lensed binary would appear as a high-mass source with the tidal deformability of a lower-mass binary. 
We demonstrate, for the first time, that this can be used as a smoking-gun evidence of lensing, or as a way to rule out the hypothesis.

Besides eliminating magnification bias, detecting lensing will be important especially for BNS systems, considering that most strong lensing cosmography studies (such as measurements of the Hubble constant, accurate tests of the speed of gravity and polarization tests) require an electromagnetic counterpart~\cite{PhysRevD.86.022004,PhysRevLett.118.091101,PhysRevD.95.063512,PhysRevLett.118.091102,2018EPJC...78..692Y}.

The article is structured as follows. In Sec.~\ref{sec:BNS_lensing}, we describe how lensing will effect the gravitational wave signal observed from a BNS merger. Sec~\ref{sec:break_degeneracy} introduces the methodology to break the degeneracy between magnification and distance measurement via the tidal deformation of a BNS. Sec~\ref{sec:highmass_binary} compares the recovery of magnification between the tidal measurement and assumed binary mass population from simulated signals. 
We then apply our methods to GW190425, finding no significant evidence to favor the lensed scenario (with a log Bayes factor $\log \mathcal{B}^L_U=-0.608^{+0.046}_{-0.021}$), and constraining the lensing magnification $\mu \leq 86.5^{+0.5}_{-11.2}$.
Finally, we provide an outlook for future
lensed gravitational-wave detections in Sec.~\ref{sec:conclusion}.

\section{Binary neutron star lensing}
\label{sec:BNS_lensing}

\begin{figure}
    \centering
    \includegraphics[width=\linewidth]{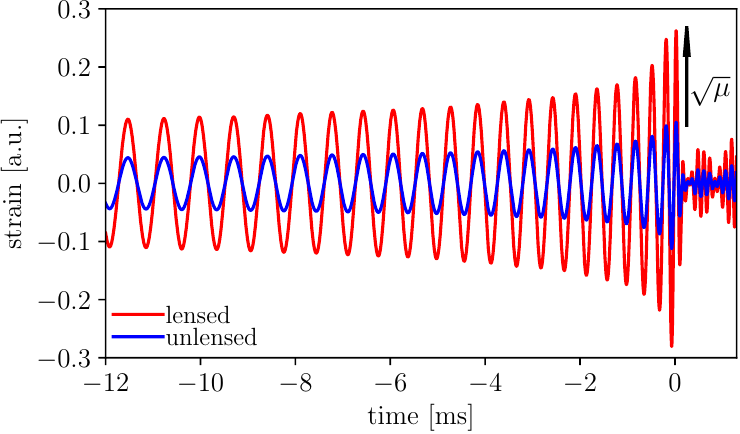}
    \caption{\textit{Illustration of amplitude magnification}: GW strain of a lensed (red) and unlensed (blue) signal from an example  binary neutron star system. The waveform shows the late inspiral and early postmerger phase and is adjusted from a numerical relativity simulation performed by the CoRe collaboration with the BAM code~\protect\cite{Dietrich:2018phi}.
    } 
    \label{fig:NR_wave_lense}
\end{figure}

The GW signal of a non-eccentric BNS coalescence is completely
described by its components' masses $m_{1,2}$, spins $\vec{s}_{1,2}$
and the supranuclear equation of state(s) governing the internal physics of both neutron
stars. There are a number of ways in which a signal emitted by a BNS
system will differ from a BBH system with the same masses and spins, due
to the presence of matter. These include the complex post-merger
signal~\cite{Bauswein:2012ya,Takami:2014zpa,Bauswein:2014qla,Bernuzzi:2015rla,Rezzolla:2016nxn,Tsang:2019esi}, the deformation of the neutron stars due to tidal
forces~\cite{Hinderer_2010,Damour:2009vw}, and the deformation of the neutron stars due to their
own rotation~\cite{Laarakkers:1997hb,Poisson:1997ha,Harry:2018hke}. Of these effects, the deformation of the
neutron star due to tidal forces provides the best measurable constraint on the internal structure and equation of state~\cite{Agathos:2015uaa,Samajdar:2019ulq}.
The tidal deformability determines the deformation of each neutron star in the gravitational field of the companion and is quantified by the parameter~\cite{Hinderer_2010,Flanagan_2008}
\begin{equation}\label{eq:tidaldeformability}
	\Lambda = \frac{2}{3}k_2\left(\frac{R}{m}\right)^5,
\end{equation}
where $k_2$, $m$, $R$ are the 2$^{\textrm{nd}}$ Love number, the mass, and the radius of the individual neutron stars, respectively. 
The tidal deformability as a function of mass can be obtained by solving the TOV equation~\cite{PhysRevD.81.123016} with a given EOS. 
These parameters depend strongly on the equation of state.

When a gravitational wave signal is lensed by intervening galaxies or galaxy clusters, the lensing magnifies the signal, increasing its amplitude without changing the signal morphology; cf. Fig.~\ref{fig:NR_wave_lense}. 
The effect is degenerate with the luminosity distance as measured from the gravitational-waves~\cite{PhysRevD.97.023012}
\begin{equation}\label{eq:luminositydistancedegeneracy}
	D^{\rm est} = \frac{D}{\sqrt{\mu}},
\end{equation}
where $D^{\rm est}$ and $D$ are the observed and true luminosity distances, respectively, and $\mu$ is the magnification induced by gravitational lensing. The measured redshift $z^{\rm est}\equiv z(D^{\rm est})$ is therefore also biased\footnote{The luminosity distance (either observed or intrinsic) is related to redshift under the assumption of standard $\Lambda$CDM cosmology~\cite{Ade:2013zuv}.}.
Redshift will cause a shift in the observed masses
of the signal according to
\begin{equation}
 m_i^{\rm est} = \frac{m_i^{\rm det}}{1+z^{\textrm{est}}}\,,
\end{equation}
where $m_i^{\rm est}$ and $m_i^{\rm det}$ is the estimated source mass and the observed detector-frame mass of each component, respectively.
Therefore in the case of a lensed source not including the lensing
magnification when characterizing the source will bias the inferred
distance, redshift and mass of the system.

Since the gravitational-wave morphology is unchanged by lensing (Fig.~\ref{fig:NR_wave_lense}), the parameters which we directly infer from the gravitational-wave phasing are unchanged~\cite{Takahashi_2003}.\footnote{Let us note that Ref.~\cite{2017arXiv170204724D} suggested that when the GW passes through a lensing saddle point, the signal morphology could exhibit a minor change. 
It was suggested that this could lead to a bias of $45 \, \deg$ in the orbital line-of-sight, and possibly minor changes in the phasing for eccentric binaries and in the higher modes of merger/ringdown. 
We have neglected such potential effects as we consider only the inspiral part where the morphology is likely to be unchanged.
However, they could be included by convolving the waveform with the complex magnification in a future study.
} 
That is, among others, the detector-frame masses $m_{1,2}^{\rm det}$ and the observed tidal deformabilities $\Lambda_{1,2}$, which are redshift independent~\cite{PhysRevLett.108.091101}, both remain unbiased. 
At leading order, the individual tidal deformabilities enter the GW phasing in a mass-weighted average $\tilde{\Lambda}$, which is given by~\cite{Flanagan_2008,Wade_2014,Favata_2014}
\begin{equation}
\begin{aligned}
    \tilde{\Lambda} &= \frac{8}{13}\bigg{(}(1+7\eta-31\eta^2)(\Lambda_1+\Lambda_2)\\
    &+ \sqrt{1 - 4\eta}(1+9\eta-11\eta^2)(\Lambda_1-\Lambda_2)\bigg{)},
\end{aligned}
\label{eq:lambdaTilde}
\end{equation}
where $\eta \equiv m_1m_2/(m_1+m_2)^2$ is the symmetric mass ratio.
Because the tidal effects can be \emph{estimated} from the masses, we will obtain two independent measurements of the tidal effects: 
First, the unbiased measurement of $\Lambda_i$ directly from the waveform phasing. 
Secondly, the \emph{estimated} $\Lambda_i^{\rm est}=\Lambda(m_i)$, from the estimate of the masses, combined with Eq.~\ref{eq:tidaldeformability}.

By making use of the above definitions, the hypothesis that the source is lensed
\begin{align}\label{eq:lensed_hypothesis}
\begin{split}
\HL:\thinspace &D = \sqrt{\mu}D^{\rm est}\,,\\
&z = z(\sqrt{\mu}D^{\rm est})\,,\\
&m_i = \frac{m_i^{\rm det}}{1+z} = m_i^{\rm est}\frac{1+z^{\rm est}}{1+z}\,,\\
&\Lambda^{\textrm{est}}_i = \Lambda(m_i) = \Lambda\left(m_i^{\rm est}\frac{1+z^{\rm est}}{1+z}\right)\,,
\end{split}
\end{align}
and, similarly, the hypothesis that the source is unlensed
\begin{align}
\begin{split}
\HU:\thinspace & m_i = m_i^{\rm est},\\
&D =D^{\rm est}\,,\\
&\Lambda^{\textrm{est}}_i = \Lambda(m_i) = \Lambda(m_i^{\rm est})\,,
\end{split}	
\end{align}
where $z(D)$ is the redshift as a function luminosity distance $D$ with a cosmological model given. 
That is, in the lensed hypothesis, the estimated masses and distances will be biased by the magnification, whereas in the unlensed one, they are their intrinsic (source-frame) quantities. 
We assume a high-magnification prior $p(\mu) \propto \mu^{-3}$ for $\mu\in[2,6000]$, which is generally a power-law near caustics~\cite{1986ApJ...310..568B}.

\begin{figure}
    \centering
    \includegraphics[width=\linewidth]{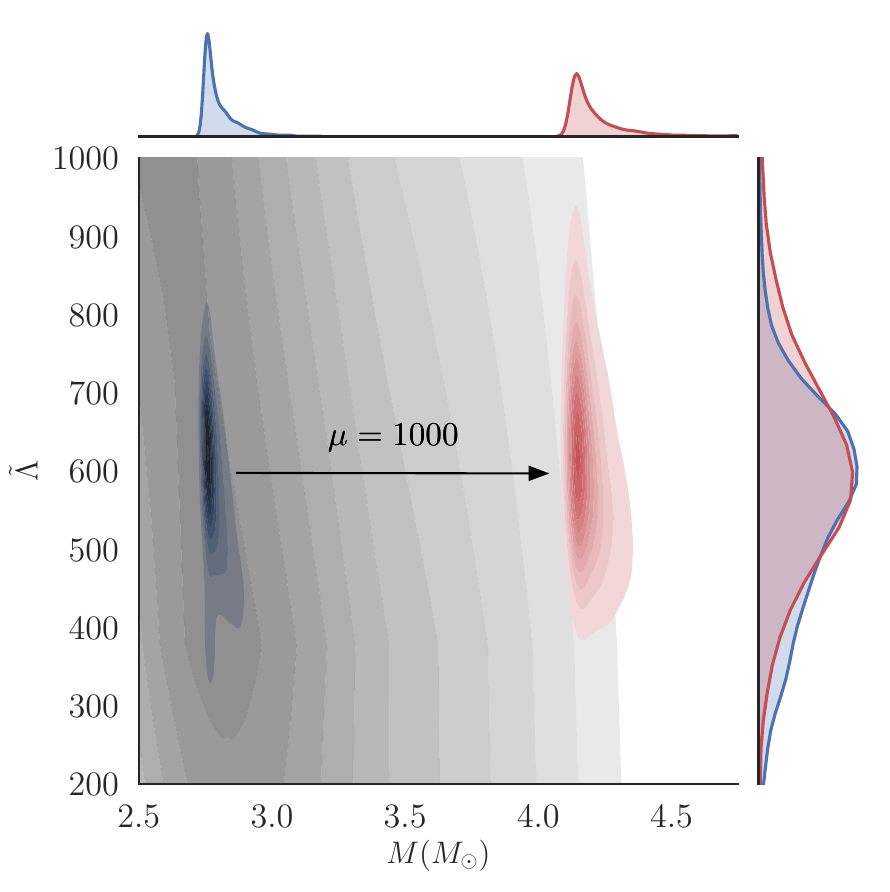}
    \caption{\textit{Effect of lensing on inferred parameters}: Corner plot of the posterior distribution of the estimated source total mass $M$ and the tidal deformability $\tilde{\Lambda}$ of the same binary neutron star merger with (\textit{red}) and without (\textit{blue}) magnification. The plot demonstrates the effect of lensing on a binary neutron star merger signal. It biases the estimated source mass to larger values
    without affecting the observed tidal deformability . The expected distribution of $\tilde{\Lambda}$-$M$ with the ENG EOS~\protect\cite{1996ApJ...469..794E} is also shown (\textit{grey}), the increase of the estimated source mass due to lensing creates tension between the expected and measured values of $\tilde{\Lambda}$-$M$.}
    \label{fig:magnificationIllustration}
\end{figure}

Consequently, the effect of the lensing magnification is to increase the observed source-frame masses, 
while the measured tidal deformability remains unchanged.
This is illustrated in Figure~\ref{fig:magnificationIllustration}, where we simulate a BNS source with a luminosity distance of $D^{\rm est}=100\textrm{Mpc}$ and source-frame masses $(1.35,1.35)$, with and without lensing magnification.

\section{Breaking the lensing degeneracy} 
\label{sec:break_degeneracy}

The tidal deformability of a BNS can be obtained in three ways: directly from the gravitational-wave phasing measurement, e.g.~\cite{Flanagan_2008,PhysRevX.9.031040,De:2018uhw}, from the observation of electromagnetic counterparts~\cite{Bauswein:2017vtn,Margalit:2017dij,Most:2018hfd,Coughlin:2018miv,Coughlin:2018fis,Radice:2018ozg}, 
or from the measured masses $m_{1,2}$ under the assumption of a given (known) EOS. 

Unfortunately, despite recent advances, the exact equation of state (EOS) governing the interior of neutron stars, i.e., 
cold matter at supranuclear densities, is still unknown. 
Information about the neutron star EOS can be obtained from nuclear physics computation, e.g., \cite{Annala:2017llu,Capano:2019eae}, from the observation of radio pulsars, e.g., \cite{Cromartie:2019kug}, 
or from the multi-messenger observation of compact binary mergers, e.g., \cite{Radice:2017lry}. 
Considering the latter, analysis of the GW signal GW170817~\cite{PhysRevLett.119.161101} disfavored
a number of theoretically allowed EOSs, which predict large tidal deformabilities and consequently large neutron star radii. 
Meanwhile, the electromagnetic observation of AT2017gfo and sGRB170817~\cite{Abbott_2017,2017ApJ...848L..33A,Chornock:2017sdf,Coulter_2017,Drout_2017,Evans_2017,Hallinan_2017,Kasliwal_2017,Murguia_Berthier_2017,Nicholl:2017ahq,Smartt_2017,Soares_Santos_2017,2017ApJ...848L..27T,Tanaka_2017,Troja_2017} disfavored EOSs with too small 
tidal deformabilities, i.e., too soft EOSs~\cite{Radice:2017lry}. 
In the future, with a growing number of multi-messenger detections of BNS mergers, and additional experiments, e.g.~NICER~\cite{10.1117/12.926396},
constraints on the allowed range of EOSs will greatly improve.

\begin{figure}
    \centering
    \includegraphics[width=\linewidth]{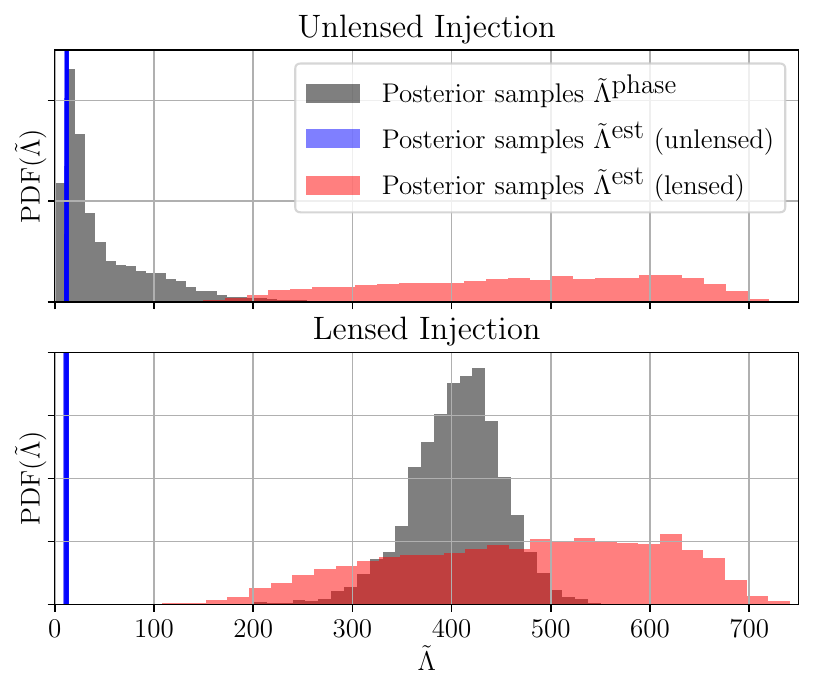}
    \caption{\textit{Effect of lensing on the estimated tidal deformability}: The posterior distribution of the tidal deformability as estimated from the gravitational-wave phasing (gray), and the binary masses under the unlensed and lensed hypothesis (blue and red, respectively). In the unlensed case (top panel), the posterior measurement, as estimated from the gravitational-wave phasing, overlaps with the unlensed prediction (blue), favoring the unlensed hypothesis. Vice versa, in the lensed case (bottom panel), the posterior measurement (gray) overlaps with the lensed prediction (red), supporting the lensed hypothesis.  The intrinsic binary masses $m_1=m_2=2.02\, \rm M_\odot$ ($m_1=m_2=1.35\, \rm M_\odot$) for the unlensed (lensed) case, while the estimated masses $m_1^{\rm est} = m_2^{\rm est} = 2.02\, \rm M_\odot$ in both cases. In this illustration, we assume SFHo equation of state~\protect\cite{Steiner_2013} and assume fixed magnification $\mu=1000$ to allow for a clear visual illustration. We show the case with variable magnification in Fig.~\ref{fig:lensingMagnification}.}
    \label{fig:magnificationIllustration2}
\end{figure}

Given an EOS, the posterior distribution of tidal deformabilities as estimated from the (observed) binary component masses under the \emph{unlensed} hypothesis is 
\begin{equation}
\begin{aligned}
    p(\Lambda^{\rm est}_i|d, \textrm{EOS}, \HU) 
    = &\int dm^{\textrm{det}}_i dz^{\rm est} \delta\left(\Lambda^{\rm est}_i - \Lambda\left(\frac{m_i^{\rm det}}{1+z^{\textrm{est}}}\right)\right)\\
    &\times p(m^{\textrm{det}}_i,z^{\rm est} |d, \HU)\,,
\end{aligned}
\end{equation}

where 
\begin{equation}
\begin{split}
    p(m_i^{\rm det}, z^{\textrm{est}}|d, \HU) = &\int dD^{\rm est} \delta(z^{\textrm{est}} - z(D^{\rm est}))\\
    &\times p(m_i^{\rm det}, D^{\rm est}|d,\HU)\,.
\end{split}
\end{equation}
The joint posterior $p(m_i^{\rm det}, D^{\rm est}|d,\HU)$ is the posterior inferred by \texttt{LALInference}. 
If the event is lensed, the lensing biases the tidal deformability under the unlensed hypothesis $p(\Lambda^{\rm est}_i|d, \textrm{EOS}, \HU)$, as predicted from the EOS, towards smaller values (as described in Sec.~\ref{sec:BNS_lensing}).

When lensing at a given magnification is taken into account, the tidal deformability estimate becomes
\begin{equation}
\begin{aligned}
    p(\Lambda^{\rm est}_i|d, \mu, \textrm{EOS}, \HL) 
    = &\int dm^{\textrm{det}}_i dz \delta\left(\Lambda^{\rm est}_i - \Lambda\left(\frac{m_i^{\rm det}}{1+z}\right)\right)\\
    &\times p(m^{\textrm{det}}_i,z |d, \mu, \HL),
\end{aligned}
\end{equation}

where 
\begin{equation}
\begin{split}
    p(m_i^{\rm det}, z|d, \mu,\HL) = &\int dD^{\rm est} \delta(z - z(\sqrt{\mu}D^{\rm est}))\\
    &\times p(m_i^{\rm det}, D^{\rm est}|d,\HL)\,.
\end{split}
\end{equation}

However, we also obtain an \emph{independent} posterior measurement of the tidal deformability $p(\Lambda^{\textrm{phase}}|d)$ directly from the gravitational-wave phasing, which is unbiased by lensing. 
By doing so, we can break the magnification-induced degeneracy by matching the two independent posterior measurements ($p(\Lambda^{\textrm{phase}}|d)$ and $p(\Lambda^{\rm est}_i|d,\mu)$) together, and rule out or confirm lensing.

\section{Discriminating between high-mass binaries and lensed binaries} \label{sec:bnslensingresults}
\label{sec:highmass_binary}

Currently known binary neutron star systems, excluding GW observations, come from Galactic observations, which consists of relatively low-mass binaries where the total mass follows roughly a normal distribution with a $2.69 \,\rm M_\odot$ mean and $0.12\, \rm M_\odot$ standard deviation~\cite{Farrow_2019}. 
If a high-mass BNS system was observed with GWs it could be considered
that it is a lensed system consistent with the Galactic population.
It would then appear as an intrinsically high-mass BNS with an apparently high tidal deformability. On the other hand, the system could belong to
a new population of high-mass BNSs. 
If such a binary was observed, it would also appear as a high-mass BNS, but with an apparently low tidal deformability.

Let us therefore show a simple illustrative example how to distinguish between these two scenarios by use of tidal measurements. 
\begin{table}
    \centering
    \begin{tabular}{l c c}
    EOS  & Lensed ($\mu=1000$) & Unlensed ($\mu=1$)\\
    \hline
    SFHo & (1.35, 432.94) & (2.02, 11.84)\\
    ENG  & (1.35, 644.66) & (2.02, 24.25)\\
    \hline
    \end{tabular}
    \caption{Summary of the source-frame mass and the tidal deformaility of the simulated binary neutron star mergers. Each cell shows the source-frame mass, tidal deformability pair $(m,\Lambda)$ of the injection under different EOS and lensing scenario.}
    \label{tab:injection_parameters}
\end{table}
For this purpose, we simulate a gravitational-wave signal from a ($m_1=m_2=1.35\,\rm M_\odot$) \emph{lensed BNS} at $\mu=1000$, consistent with the Galactic double neutron star population, at an observed distance of $100$ Mpc, assuming LIGO/Virgo detector network at design sensitivity, and described by the SFHo\cite{Steiner_2013} and ENG\cite{1996ApJ...469..794E} EOSs \footnote{These particular EOSs are chosen since they are broadly in agreement with joint-constraint derived from GW170817 and AT2017gfo, e.g.,~\cite{Radice:2017lry,Radice:2018ozg,Coughlin:2018fis}}.

For our analysis, we employ the standard LVC-developed nested sampling framework,  \textsc{LALInference} (see Appendix~\ref{app:PE} and Ref.~\cite{Veitch:2014wba,lalsuite} for details). 
We recover the tidal deformability from the gravitational-wave phasing (Method-I) and from the EOS and masses (Method-II) (see Fig.~\ref{fig:magnificationIllustration2}, bottom panel, gray and blue bins, respectively)\footnote{We assume the SFHo EOS~\cite{Steiner_2013}, for simplicity.}.
The results disagree with each other, ruling out the unlensed hypothesis.
Then, we assume that the event is lensed at a magnification of $\mu=1000$, and repeat the measurement (Fig.~\ref{fig:magnificationIllustration2}, bottom panel, gray and red bins, respectively). 
The posteriors overlap, supporting the lensing hypothesis.

We then demonstrate the same test for a ($m_1=m_2=2\,\rm M_\odot$) \emph{unlensed but high-mass BNS} at an observed distance $D^{\rm est}$ of $100$ Mpc. 
In this case, the tidal deformability from the gravitational-wave phasing and from the EOS/masses overlap (see Fig.~\ref{fig:magnificationIllustration2}, top panel), favoring the unlensed hypothesis.
Thus, the test can be used to discriminate between intrinsically high-mass BNSs and lensed BNSs.
Note that here, for the sake of illustrating the method, we have fixed the magnification; we show the more general case with variable magnification below. 
\\

\begin{figure}
    \centering
    \includegraphics[width=\linewidth]{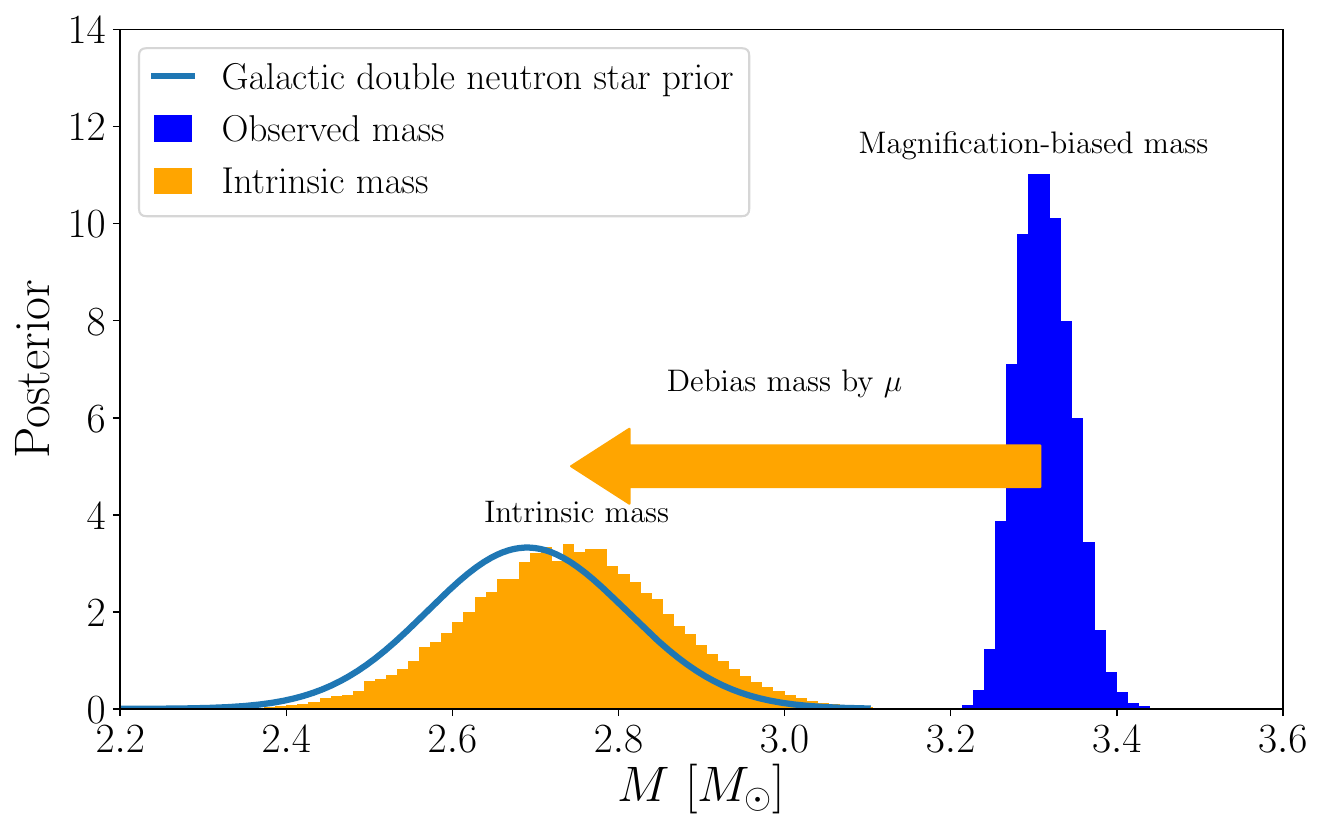}
    \caption{Illustration of the magnification posterior computation for GW190425 from the binary masses, assuming the galactic double neutron star population as the mass prior. 
    The figure shows the observed total binary mass for GW190425 (orange), the galactic double neutron star population prior (green), and the mass distribution if the event was lensed (blue). 
    By re-weighting the unlensed posterior, we can estimate the lensing magnification $\mu$ under the lensing hypothesis. 
    Note that we can also estimate the lensing magnification independently of the galactic double neutron star prior using tidal effects, as illustrated in Fig.~\ref{fig:magnificationIllustration2}.
    }
    \label{fig:MagnificationPosteriorIllustration}
\end{figure}

Let us now consider the more general case with arbitrary magnification, instead of fixed magnification. 
Given a source population (which we assume to be the galactic double neutron star population), we can estimate the lensing magnification $p(\mu|d, \HLDNS)$ where we have explicitly defined the hypothesis $\HLDNS$ to refer to the magnification estimate from the binary \emph{masses} (see Appendix~\ref{app:methods} for the detailed derivation). I.e., the mass prior $p(M|\HLDNS)$ is the one for galactic double neutron stars (a normal distribution with a $2.69 \,\rm M_\odot$ mean and $0.12\, \rm M_\odot$ standard deviation~\cite{Farrow_2019}), but we make no explicit constraint on the tidal measurements.
This is done by \emph{unbiasing} the GW measurement such that it is consistent with the expected source population (see Fig.~\ref{fig:MagnificationPosteriorIllustration}, for an illustration of the process for GW190425). 

Alternatively, we can estimate the magnification $\mu$ by combining the estimated tidal deformability with the directly measured one (see Appendix~\ref{app:methods})
\begin{equation}
    p(\mu | d, {\rm EOS}, \HLT) \propto \left\langle \frac{p(\tilde{\Lambda}^{\textrm{phase}}  | d,\HU)}{p(\tilde{\Lambda}^{\textrm{phase}}| q, \HU)}\mathcal{W}_{\textrm {EOS}} \right\rangle_{\tilde{\Lambda}^{\textrm{phase}} = \tilde{\Lambda}^{\textrm{est}} },
\end{equation}
where $p(\tilde{\Lambda}^{\textrm{phase}}|d,\HU)$ is the posterior distribution of the measured tidal deformability under the unlensed hypothesis, $\tilde{\Lambda}^{\textrm{est}}$ is the estimated tidal deformability with a given magnification and EOS, and $\langle \cdots \rangle$ refers to an average over the mass and distance posterior samples.
The weight $\mathcal{W}_{\rm EOS}$ is given by
\begin{equation}
\begin{aligned}
    \mathcal{W}_{\rm EOS} &= \frac{p(m_1^{\rm est}, m_2^{\rm est}| D^{\rm est}, \mu, \HLT, \textrm{EOS} )p(D^{\rm est}|\mu, \HL)}{p(m_1^{\rm est}, m_2^{\rm est}|D^{\rm est}, \HU)p(D^{\rm est}| \HU)}\\
    &\,\,\,\,\,\,\,\,\times p(\mu|\HL)\,.
\end{aligned}
\end{equation}
Here $\HLT$ refers to the lensed hypothesis that additionally enforces 
\begin{equation}
    p(\Lambda,\Lambda^{\textrm{est}}|\HLT)=p(\Lambda|\HL)p(\Lambda^{\textrm {est}}|\HL)\delta(\Lambda^{\textrm{est}}-\Lambda)
\end{equation} in the prior. The mass prior under $\HLT$ hypothesis is taken to be a flat prior between 0.5$M_\odot$ and the maximum mass allowed by the EOS. 

We can calculate the evidence for the lensed hypothesis $Z_{L}$ and the unlensed hypothesis $Z_{U}$ by
\begin{equation}
\begin{aligned}
    Z_L &= \int d\mu p(d|\mu,\textrm{EOS},\HLT) p(\mu)\\
    Z_U &= p(d|\mu=1,\textrm{EOS}).
\end{aligned}
\end{equation}

The log Bayes factor $\log\mathcal{B}^{L}_{U}$ is defined as the log of the ratio between the two evidence, therefore $\log\mathcal{B}^{L}_{U} \equiv \log(Z_L/Z_U)$. A positive $\log\mathcal{B}^{L}_{U}$ shows that the lensed hypothesis is more plausible than the unlensed hypothesis. For the analysis, we consider a range of EOSs, which are SFHo, ENG and MPA1. These EOSs show agreement with the joint-constraint obtained with GW170817 and AT2017gfo~\cite{Radice:2017lry}.

Since the tidal deformability measurement is not biased by lensing, we expect this secondary measurement of the magnification to be independent of any assumptions on the source population (i.e., it is completely unbiased). 
Therefore, we expect the magnification to be low for unlensed binaries, and high for lensed binaries. 

\begin{figure*}
    \centering
    \includegraphics[width=\columnwidth]{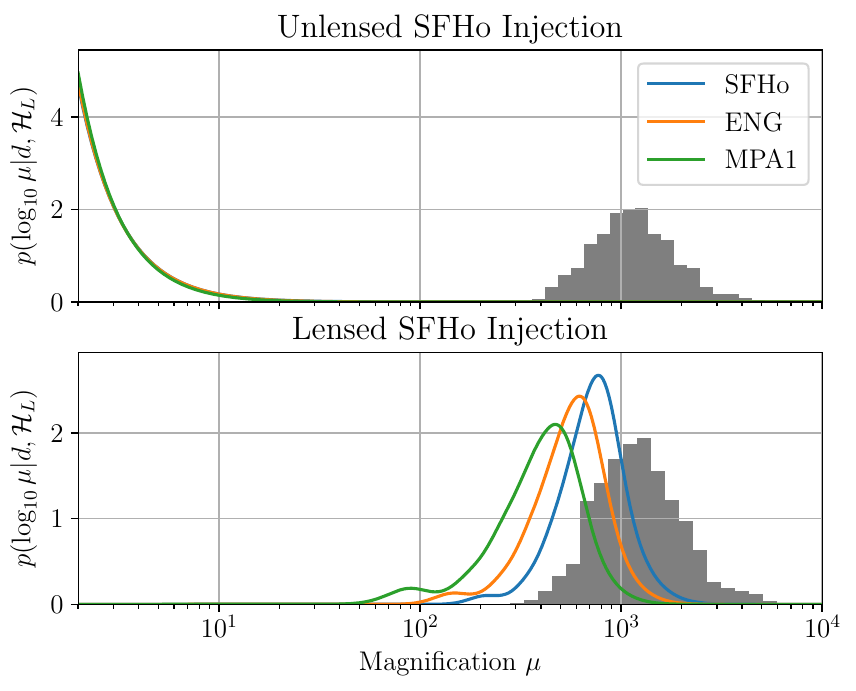} \includegraphics[width=\columnwidth]{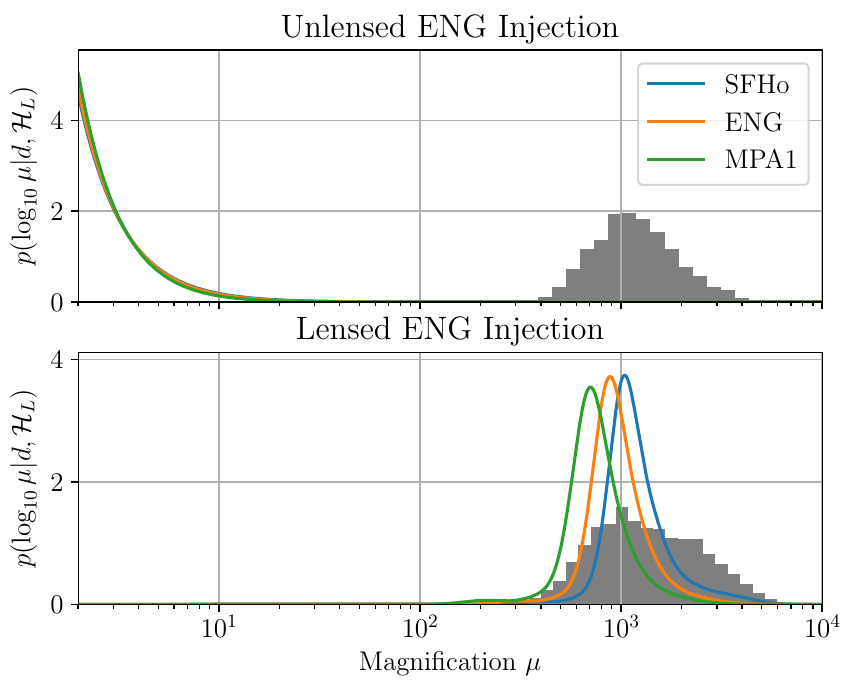}
    \caption{Posterior distribution of magnifications inferred with posteriors of component masses and luminosity distance (gray bins) and that with posteriors of component masses and tidal deformability for given EOSs (colored line) with various injections. 
    We show four different injections: Unlensed SFHo (top left), unlensed ENG (top right), lensed SFHo (bottom left) and lensed ENG (bottom right) injection.
    The posterior of the magnification $\mu$ inferred from the masses and from the tidal deformabilities are giving consistent results for lensed injections. Meanwhile, there exists tension between the posteriors recovered by the two means for unlensed injection. 
    The injected BNS masses are ($m_1=m_2=1.35\, \rm M_\odot$) and ($m_1=m_2=2.02 \, \rm M_\odot$) for the lensed and unlensed binaries, respectively. 
    The binary neutron star is at an observed luminosity distance of $D^{\rm est}=100\,\rm Mpc$, with a signal-to-noise ratio of 31.}
    \label{fig:lensingMagnification}
\end{figure*}

Fig.~\ref{fig:lensingMagnification} shows the magnification posteriors evaluated via the two methods above, for both the lensed and unlensed injections with different EOSs (Table~\ref{tab:injection_parameters}).
We observed that the required magnifications $p(\mu|d,\HLDNS)$, as evaluated from the galactic double neutron star population, are in the $\mu\sim \mathcal{O}(100) - \mathcal{O}(1000)$ range for both the lensed and unlensed injections (Fig.~\ref{fig:lensingMagnification}, gray bins). 
Meanwhile, the magnifications as estimated from the unbiased tidal deformabilities are different for the two scenarios, favoring the unlensed case for the unlensed injection, and lensed case for the lensed injection (solid lines, for the SFHo, ENG and MPA1 EOSs).
Most notably, we find that the two magnification estimates disagree in the unlensed case, ruling out the lensed hypothesis at a log Bayes factor $\log\mathcal{B}^L_U$ of $-2.72(-2.68)$, $-2.75(-2.71)$ and $-2.82(-2.83)$ for SFHo, ENG and MPA1, respectively, for SFHo(ENG) injection. And agree in the lensed case, confirming the hypothesis at a log Bayes factor $\log\mathcal{B}^L_U$ of $38.5(36.26)$, $31.2(32.63)$ and $20.9(26.75)$ for SFHo, ENG and MPA1, respectively, for SFHo(ENG) injection. For the unlensed case, the posterior of the magnification $\mu$ rails against the prior instead of peaking at the true value (therefore $\mu=1$), which result in the log Bayes factor $\log\mathcal{B}^L_U$ to be in different magnitude for the lensed and unlensed injections.
As a supplementary analysis, we also performed the estimate on an injection set with a magnification of 100, finding that we can still disfavor lensing for the high-mass binary, but that we are unable to confirm lensing in this case (Appendix~\ref{app:mu100}).

\section{Beyond mock data: Discussion}

Our work demonstrates a robust methodology to rule out or confirm the gravitational lensing hypothesis for BNS mergers. 
The methodology can be used to rule out lensing for intrinsically high mass BNS events, or confirm it for the galactic double neutron star population. 
The mock data was produced for two different lensed and unlensed scenario, employing the SFHo and ENG equations of state consistent with the current GW and EM observations~\cite{Radice:2017lry}. 
It is natural to wonder if the analysis could already rule out or confirm lensing for the high-mass binary neutron star event GW190425, and if not, what is required of a realistic detection to be able to make this distinction. 

\begin{figure}
    \centering
    \includegraphics[width=\linewidth]{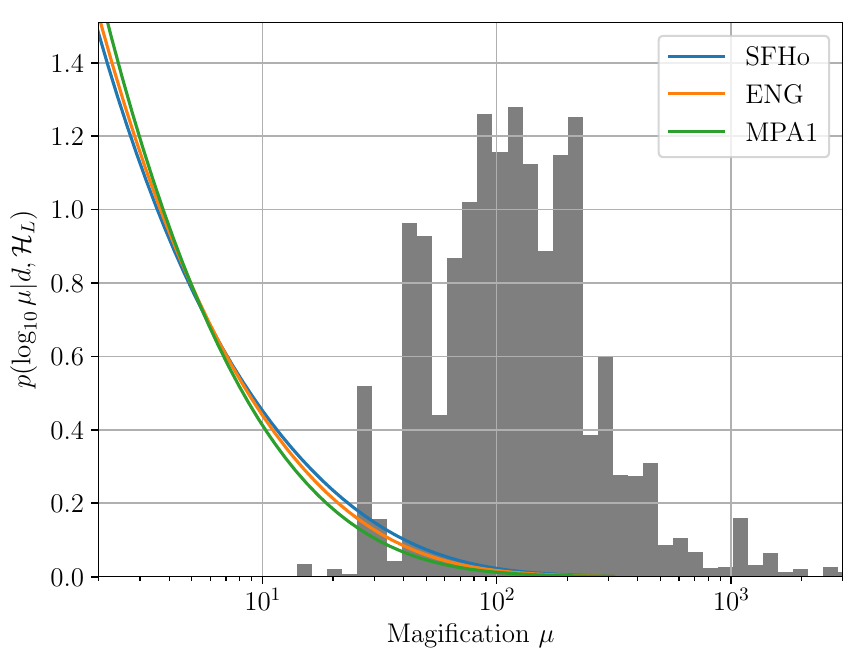}
    \caption{Posterior distribution of magnifications inferred with the galactic  double  neutron  star  population assumed (gray bins) and posterior  distribution  of  magnifications inferred  with different EOSs assumed (colored lines). As the  lensed and unlensed magnification estimates overlap, allowing no clear constraints on the lens hypothesis. We deduce that the magnification $\mu$ is less than $87.0$, $86.5$ and $75.3$ for SFHo, ENG and MPA1, respectively, at a 99$\%$ confident.}
    \label{fig:lensingMagnificationGW190425}
\end{figure}

We evaluate the magnification posterior using both the mass estimate and the tidal deformability measurement (as in Sec.~\ref{sec:bnslensingresults}) for GW190425\footnote{The parameter estimation samples released in \cite{GW190425_open_data} is used.}, but find that both the lensed and unlensed magnification estimates overlap, allowing no clear constraints on the lens hypothesis (Fig.~\ref{fig:lensingMagnificationGW190425}). 
However, we note that binary neutron star lensing is very unlikely within LIGO/Virgo at current sensitivity. 
Thus, in the absence of evidence, it is plausible that the event is not lensed.
The log Bayes factor for the lensed hypothesis against unlensed hypothesis are shown in Table~\ref{tab:GW190425logB} for a selected set of EOSs.
\begin{table}
    \centering
    \begin{tabular}{c c}
         EOS & $\log\mathcal{B}^L_U$ \\
         \hline
         SFHo & -0.610 \\
         ENG & -0.646 \\
         MPA1 & -0.715\\
         \hline
    \end{tabular}
    \caption{The log Bayes factor for the lensed hypothesis against unlensed hypothesis of GW190425 with various EOSs given.}
    \label{tab:GW190425logB}
\end{table}
We deduce that the magnification $\mu$ is less than $87.0$, $86.5$ and $75.3$ for SFHo, ENG and MPA1, respectively, at a $99\%$ confidence level. 

Had the event been observed at design sensitivity, and in the full detector network (LIGO Hanford/Livingston and Virgo), the network SNR would have been $\sim 23$, which is much closer to the signal strengths which we used in our mock data simulations (SNR $\sim$ $30$). 
Therefore, while we can not set very stringent constraints on lensing for the GW190425 event, a similar event at a lower distance detected by  LIGO/Virgo or the same event with more sensitive instruments, in the future, might allow us to probe the lensing hypothesis.

Moreover,  we note that the lensing hypothesis can be ruled out more easily for higher mass events. 
The total mass of the GW190425 event was $3.4_{-0.1}^{+0.3} \, \rm M_\odot$, which would already necessiate fairly large magnifications if it were lensed (Fig.~\ref{fig:lensingMagnificationGW190425}). 
If the BNS population which produced GW190425 consists of higher mass BNS events, we will likely be able to set better constraints.

If the event is indeed lensed at a high magnification, then our method can be used to confirm that the event is lensed. 
It is currently unlikely that we will detect binary neutron star lensing within LIGO/Virgo. 
However, with future third-generation detectors such as the Einstein Telescope, lensed detections could be in the hundreds~\cite{2014JCAP...10..080B,2015JCAP...12..006D}. 
We could discover these events at a much higher SNR than, allowing for more robust constraints than presented here. 

As we observe more BNS events, we will be able to set more stringent constraints on the EOS of neutron stars due to the combination/stacking of multiple gravitational wave sources~\cite{DelPozzo:2013ala,Agathos:2015uaa} and their potential EM counterparts. 
Therefore, our estimate of the expected tidal deformabilities will improve, which in turn will allow for improved tests of the BNS lensing. 
Future studies employing populations of events will answer the above questions more definitively.

\section{Conclusions}
\label{sec:conclusion}
If a GW from a BNS event is lensed, a combined measurement of the tidal effects and the binary masses of BNSs could be used to rule out or confirm the lensing hypothesis robustly. 
This test could be used to rule out lensing for intrinsically high-mass BNSs, similar to the recent GW190425 event. 
Lensed BNSs are one of the GW sources that can be gravitationally lensed and  produce an electromagnetic counterpart. This makes them to an attractive target for multi-messenger studies. 
Indeed, lensed BNSs might allow for measurements of the Hubble constant~\cite{Liao_2017}, accurate tests of the speed of gravity~\cite{PhysRevLett.118.091101,PhysRevLett.118.091102}, various cosmography studies~\cite{Smith_2019b}, and polarization tests~\cite{PhysRevD.86.022004}.
Since our test could also be used to robustly confirm BNS lensing, it is expected to find several use-cases in these novel strong lensing avenues that utilize EM counterparts.

\section*{Acknowledgements}
  We thank Tjonnie Li, David Keitel, Ken Ng, Ajit Mehta, K. Haris, Chris Van Den Broeck, Christopher Berry and Shasvath Kapadia for their comments and suggestions. 
  PTHP and OAH are supported by the research program of the Netherlands Organization for Scientific Research (NWO).
  TD acknowledges support by the European Union’s Horizon 
  2020 research and innovation program under grant
  agreement No 749145, BNSmergers. We
are grateful for the computing resources provided by the
LIGO-Caltech Computing Cluster where our analysis
was carried out.

\bibliographystyle{mnras}
\bibliography{main}

\begin{thebibliography}{}
\makeatletter
\relax
\def\mn@urlcharsother{\let\do\@makeother \do\$\do\&\do\#\do\^\do\_\do\%\do\~}
\def\mn@doi{\begingroup\mn@urlcharsother \@ifnextchar [ {\mn@doi@}
  {\mn@doi@[]}}
\def\mn@doi@[#1]#2{\def\@tempa{#1}\ifx\@tempa\@empty \href
  {http://dx.doi.org/#2} {doi:#2}\else \href {http://dx.doi.org/#2} {#1}\fi
  \endgroup}
\def\mn@eprint#1#2{\mn@eprint@#1:#2::\@nil}
\def\mn@eprint@arXiv#1{\href {http://arxiv.org/abs/#1} {{\tt arXiv:#1}}}
\def\mn@eprint@dblp#1{\href {http://dblp.uni-trier.de/rec/bibtex/#1.xml}
  {dblp:#1}}
\def\mn@eprint@#1:#2:#3:#4\@nil{\def\@tempa {#1}\def\@tempb {#2}\def\@tempc
  {#3}\ifx \@tempc \@empty \let \@tempc \@tempb \let \@tempb \@tempa \fi \ifx
  \@tempb \@empty \def\@tempb {arXiv}\fi \@ifundefined
  {mn@eprint@\@tempb}{\@tempb:\@tempc}{\expandafter \expandafter \csname
  mn@eprint@\@tempb\endcsname \expandafter{\@tempc}}}

\bibitem[\protect\citeauthoryear{Abbott et~al.}{Abbott
  et~al.}{2015}]{AdvLIGO2015}
Abbott B.~P.,  et~al., 2015, \mn@doi [Classical and Quantum Gravity]
  {10.1088/0264-9381/32/7/074001}, 32, 074001

\bibitem[\protect\citeauthoryear{Abbott et~al.}{Abbott
  et~al.}{2017a}]{PhysRevLett.119.161101}
Abbott B.~P.,  et~al., 2017a, \mn@doi [Phys. Rev. Lett.]
  {10.1103/PhysRevLett.119.161101}, 119, 161101

\bibitem[\protect\citeauthoryear{Abbott et~al.,}{Abbott
  et~al.}{2017b}]{Abbott_2017}
Abbott B.~P.,  et~al., 2017b, \mn@doi [The Astrophysical Journal]
  {10.3847/2041-8213/aa91c9}, 848, L12

\bibitem[\protect\citeauthoryear{Abbott et~al.}{Abbott
  et~al.}{2019}]{PhysRevX.9.031040}
Abbott B.~P.,  et~al., 2019, \mn@doi [Phys. Rev.] {10.1103/PhysRevX.9.031040},
  X9, 031040

\bibitem[\protect\citeauthoryear{Abbott et~al.}{Abbott
  et~al.}{2020}]{collaboration2020gw190425}
Abbott B.~P.,  et~al., 2020, {GW190425: Observation of a Compact Binary
  Coalescence with Total Mass $\sim 3.4 M_{\odot}$} (\mn@eprint {arXiv}
  {2001.01761})

\bibitem[\protect\citeauthoryear{Acernese et~al.}{Acernese
  et~al.}{2014}]{Acernese_2014}
Acernese F.,  et~al., 2014, \mn@doi [Classical and Quantum Gravity]
  {10.1088/0264-9381/32/2/024001}, 32, 024001

\bibitem[\protect\citeauthoryear{Ade et~al.}{Ade et~al.}{2014}]{Ade:2013zuv}
Ade P. A.~R.,  et~al., 2014, \mn@doi [Astron. Astrophys.]
  {10.1051/0004-6361/201321591}, 571, A16

\bibitem[\protect\citeauthoryear{Agathos, Meidam, Del~Pozzo, Li, Tompitak,
  Veitch, Vitale  \& Van Den~Broeck}{Agathos et~al.}{2015}]{Agathos:2015uaa}
Agathos M.,  Meidam J.,  Del~Pozzo W.,  Li T. G.~F.,  Tompitak M.,  Veitch J.,
  Vitale S.,   Van Den~Broeck C.,  2015, \mn@doi [Phys. Rev.]
  {10.1103/PhysRevD.92.023012}, D92, 023012

\bibitem[\protect\citeauthoryear{Akutsu et~al.,}{Akutsu
  et~al.}{2018}]{10.1093/ptep/ptx180}
Akutsu T.,  et~al., 2018, \mn@doi [Progress of Theoretical and Experimental
  Physics] {10.1093/ptep/ptx180}, 2018

\bibitem[\protect\citeauthoryear{Annala, Gorda, Kurkela  \& Vuorinen}{Annala
  et~al.}{2018}]{Annala:2017llu}
Annala E.,  Gorda T.,  Kurkela A.,   Vuorinen A.,  2018, \mn@doi [Phys. Rev.
  Lett.] {10.1103/PhysRevLett.120.172703}, 120, 172703

\bibitem[\protect\citeauthoryear{{Arcavi} et~al.,}{{Arcavi}
  et~al.}{2017}]{2017ApJ...848L..33A}
{Arcavi} I.,  et~al., 2017, \mn@doi [\apjl] {10.3847/2041-8213/aa910f}, \href
  {https://ui.adsabs.harvard.edu/abs/2017ApJ...848L..33A} {848, L33}

\bibitem[\protect\citeauthoryear{Aso, Michimura, Somiya, Ando, Miyakawa,
  Sekiguchi, Tatsumi  \& Yamamoto}{Aso et~al.}{2013}]{PhysRevD.88.043007}
Aso Y.,  Michimura Y.,  Somiya K.,  Ando M.,  Miyakawa O.,  Sekiguchi T.,
  Tatsumi D.,   Yamamoto H.,  2013, \mn@doi [Phys. Rev. D]
  {10.1103/PhysRevD.88.043007}, 88, 043007

\bibitem[\protect\citeauthoryear{Baker \& Trodden}{Baker \&
  Trodden}{2017}]{PhysRevD.95.063512}
Baker T.,  Trodden M.,  2017, \mn@doi [Phys. Rev. D]
  {10.1103/PhysRevD.95.063512}, 95, 063512

\bibitem[\protect\citeauthoryear{Bauswein, Janka, Hebeler  \& Schwenk}{Bauswein
  et~al.}{2012}]{Bauswein:2012ya}
Bauswein A.,  Janka H.,  Hebeler K.,   Schwenk A.,  2012, \mn@doi [Phys.Rev.D]
  {10.1103/PhysRevD.86.063001}, 86, 063001

\bibitem[\protect\citeauthoryear{Bauswein, Stergioulas  \& Janka}{Bauswein
  et~al.}{2014}]{Bauswein:2014qla}
Bauswein A.,  Stergioulas N.,   Janka H.-T.,  2014, \mn@doi [Phys.Rev.D]
  {10.1103/PhysRevD.90.023002}, 90, 023002

\bibitem[\protect\citeauthoryear{Bauswein, Just, Janka  \&
  Stergioulas}{Bauswein et~al.}{2017}]{Bauswein:2017vtn}
Bauswein A.,  Just O.,  Janka H.-T.,   Stergioulas N.,  2017, \mn@doi
  [Astrophys. J.] {10.3847/2041-8213/aa9994}, 850, L34

\bibitem[\protect\citeauthoryear{Bernuzzi, Dietrich  \& Nagar}{Bernuzzi
  et~al.}{2015}]{Bernuzzi:2015rla}
Bernuzzi S.,  Dietrich T.,   Nagar A.,  2015, \mn@doi [Phys.Rev.Lett.]
  {10.1103/PhysRevLett.115.091101}, 115, 091101

\bibitem[\protect\citeauthoryear{{Biesiada}, {Ding}, {Pi{\'o}rkowska}  \&
  {Zhu}}{{Biesiada} et~al.}{2014}]{2014JCAP...10..080B}
{Biesiada} M.,  {Ding} X.,  {Pi{\'o}rkowska} A.,   {Zhu} Z.-H.,  2014, \mn@doi
  [\jcap] {10.1088/1475-7516/2014/10/080}, \href
  {https://ui.adsabs.harvard.edu/abs/2014JCAP...10..080B} {2014, 080}

\bibitem[\protect\citeauthoryear{{Blandford} \& {Narayan}}{{Blandford} \&
  {Narayan}}{1986}]{1986ApJ...310..568B}
{Blandford} R.,  {Narayan} R.,  1986, \mn@doi [\apj] {10.1086/164709}, \href
  {https://ui.adsabs.harvard.edu/abs/1986ApJ...310..568B} {310, 568}

\bibitem[\protect\citeauthoryear{Bliokh \& Minakov}{Bliokh \&
  Minakov}{1975}]{bliokh1975diffraction}
Bliokh P.,  Minakov A.,  1975, Astrophysics and Space Science, 34, L7

\bibitem[\protect\citeauthoryear{Bontz \& Haugan}{Bontz \&
  Haugan}{1981}]{bontz1981diffraction}
Bontz R.~J.,  Haugan M.~P.,  1981, Astrophysics and Space Science, 78, 199

\bibitem[\protect\citeauthoryear{Broadhurst, Diego  \& III}{Broadhurst
  et~al.}{2018}]{broadhurst2018reinterpreting}
Broadhurst T.,  Diego J.~M.,   III G.~S.,  2018, Reinterpreting Low Frequency
  LIGO/Virgo Events as Magnified Stellar-Mass Black Holes at Cosmological
  Distances (\mn@eprint {arXiv} {1802.05273})

\bibitem[\protect\citeauthoryear{Capano et~al.,}{Capano
  et~al.}{2019}]{Capano:2019eae}
Capano C.~D.,  et~al., 2019

\bibitem[\protect\citeauthoryear{Chatziioannou, Yunes  \&
  Cornish}{Chatziioannou et~al.}{2012}]{PhysRevD.86.022004}
Chatziioannou K.,  Yunes N.,   Cornish N.,  2012, \mn@doi [Phys. Rev. D]
  {10.1103/PhysRevD.86.022004}, 86, 022004

\bibitem[\protect\citeauthoryear{Chornock et~al.}{Chornock
  et~al.}{2017}]{Chornock:2017sdf}
Chornock R.,  et~al., 2017, \mn@doi [Astrophys. J.] {10.3847/2041-8213/aa905c},
  848, L19

\bibitem[\protect\citeauthoryear{Christian, Vitale  \& Loeb}{Christian
  et~al.}{2018}]{PhysRevD.98.103022}
Christian P.,  Vitale S.,   Loeb A.,  2018, \mn@doi [Phys. Rev. D]
  {10.1103/PhysRevD.98.103022}, 98, 103022

\bibitem[\protect\citeauthoryear{Collaboration, the Virgo~Collaboration  \& the
  KAGRA~Collaboration}{Collaboration et~al.}{2013}]{collaboration2013prospects}
Collaboration T. L.~S.,  the Virgo~Collaboration  the KAGRA~Collaboration 2013,
  Prospects for Observing and Localizing Gravitational-Wave Transients with
  Advanced LIGO, Advanced Virgo and KAGRA (\mn@eprint {arXiv} {1304.0670})

\bibitem[\protect\citeauthoryear{Collett \& Bacon}{Collett \&
  Bacon}{2017}]{PhysRevLett.118.091101}
Collett T.~E.,  Bacon D.,  2017, \mn@doi [Phys. Rev. Lett.]
  {10.1103/PhysRevLett.118.091101}, 118, 091101

\bibitem[\protect\citeauthoryear{Contigiani}{Contigiani}{2020}]{Contigiani:2020yyc}
Contigiani O.,  2020, \mn@doi [Monthly Notices of the Royal Astronomical
  Society] {10.1093/mnras/staa026}, 492, 3359

\bibitem[\protect\citeauthoryear{Coughlin et~al.,}{Coughlin
  et~al.}{2018}]{Coughlin:2018miv}
Coughlin M.~W.,  et~al., 2018, \mn@doi [Monthly Notices of the Royal
  Astronomical Society] {10.1093/mnras/sty2174}, 480, 3871

\bibitem[\protect\citeauthoryear{Coughlin, Dietrich, Margalit  \&
  Metzger}{Coughlin et~al.}{2019}]{Coughlin:2018fis}
Coughlin M.~W.,  Dietrich T.,  Margalit B.,   Metzger B.~D.,  2019, \mn@doi
  [Monthly Notices of the Royal Astronomical Society: Letters]
  {10.1093/mnrasl/slz133}, 489, L91

\bibitem[\protect\citeauthoryear{Coulter et~al.,}{Coulter
  et~al.}{2017}]{Coulter_2017}
Coulter D.~A.,  et~al., 2017, \mn@doi [Science] {10.1126/science.aap9811}, 358,
  1556–1558

\bibitem[\protect\citeauthoryear{Cromartie et~al.}{Cromartie
  et~al.}{2019}]{Cromartie:2019kug}
Cromartie H.~T.,  et~al., 2019, \mn@doi [Nat. Astron.]
  {10.1038/s41550-019-0880-2}, 4, 72

\bibitem[\protect\citeauthoryear{{Dai} \& {Venumadhav}}{{Dai} \&
  {Venumadhav}}{2017}]{2017arXiv170204724D}
{Dai} L.,  {Venumadhav} T.,  2017, arXiv e-prints, \href
  {https://ui.adsabs.harvard.edu/abs/2017arXiv170204724D} {p. arXiv:1702.04724}

\bibitem[\protect\citeauthoryear{Dai, Venumadhav  \& Sigurdson}{Dai
  et~al.}{2017}]{PhysRevD.95.044011}
Dai L.,  Venumadhav T.,   Sigurdson K.,  2017, \mn@doi [Phys. Rev. D]
  {10.1103/PhysRevD.95.044011}, 95, 044011

\bibitem[\protect\citeauthoryear{{Dai}, {Li}, {Zackay}, {Mao}  \& {Lu}}{{Dai}
  et~al.}{2018}]{2018PhRvD..98j4029D}
{Dai} L.,  {Li} S.-S.,  {Zackay} B.,  {Mao} S.,   {Lu} Y.,  2018, \mn@doi
  [\prd] {10.1103/PhysRevD.98.104029}, \href
  {https://ui.adsabs.harvard.edu/abs/2018PhRvD..98j4029D} {98, 104029}

\bibitem[\protect\citeauthoryear{Damour \& Nagar}{Damour \&
  Nagar}{2009}]{Damour:2009vw}
Damour T.,  Nagar A.,  2009, \mn@doi [Phys.Rev.D] {10.1103/PhysRevD.80.084035},
  80, 084035

\bibitem[\protect\citeauthoryear{De, Finstad, Lattimer, Brown, Berger  \&
  Biwer}{De et~al.}{2018}]{De:2018uhw}
De S.,  Finstad D.,  Lattimer J.~M.,  Brown D.~A.,  Berger E.,   Biwer C.~M.,
  2018, \mn@doi [Phys. Rev. Lett.] {10.1103/PhysRevLett.121.259902,
  10.1103/PhysRevLett.121.091102}, 121, 091102

\bibitem[\protect\citeauthoryear{Deguchi \& Watson}{Deguchi \&
  Watson}{1986}]{deguchi1986diffraction}
Deguchi S.,  Watson W.,  1986, The Astrophysical Journal, 307, 30

\bibitem[\protect\citeauthoryear{Del~Pozzo, Li, Agathos, Van Den~Broeck  \&
  Vitale}{Del~Pozzo et~al.}{2013}]{DelPozzo:2013ala}
Del~Pozzo W.,  Li T. G.~F.,  Agathos M.,  Van Den~Broeck C.,   Vitale S.,
  2013, \mn@doi [Phys. Rev. Lett.] {10.1103/PhysRevLett.111.071101}, 111,
  071101

\bibitem[\protect\citeauthoryear{Dietrich et~al.,}{Dietrich
  et~al.}{2018}]{Dietrich:2018phi}
Dietrich T.,  et~al., 2018, \mn@doi [Classical and Quantum Gravity]
  {10.1088/1361-6382/aaebc0}, 35, 24LT01

\bibitem[\protect\citeauthoryear{{Ding}, {Biesiada}  \& {Zhu}}{{Ding}
  et~al.}{2015}]{2015JCAP...12..006D}
{Ding} X.,  {Biesiada} M.,   {Zhu} Z.-H.,  2015, \mn@doi [\jcap]
  {10.1088/1475-7516/2015/12/006}, \href
  {https://ui.adsabs.harvard.edu/abs/2015JCAP...12..006D} {2015, 006}

\bibitem[\protect\citeauthoryear{Drout et~al.,}{Drout
  et~al.}{2017}]{Drout_2017}
Drout M.~R.,  et~al., 2017, \mn@doi [Science] {10.1126/science.aaq0049}, 358,
  1570–1574

\bibitem[\protect\citeauthoryear{{Engvik}, {Osnes}, {Hjorth-Jensen}, {Bao}  \&
  {Ostgaard}}{{Engvik} et~al.}{1996}]{1996ApJ...469..794E}
{Engvik} L.,  {Osnes} E.,  {Hjorth-Jensen} M.,  {Bao} G.,   {Ostgaard} E.,
  1996, \mn@doi [\apj] {10.1086/177827}, \href
  {https://ui.adsabs.harvard.edu/abs/1996ApJ...469..794E} {469, 794}

\bibitem[\protect\citeauthoryear{Evans et~al.,}{Evans
  et~al.}{2017}]{Evans_2017}
Evans P.~A.,  et~al., 2017, \mn@doi [Science] {10.1126/science.aap9580}, 358,
  1565–1570

\bibitem[\protect\citeauthoryear{Fan, Liao, Biesiada, Pi\'orkowska-Kurpas  \&
  Zhu}{Fan et~al.}{2017}]{PhysRevLett.118.091102}
Fan X.-L.,  Liao K.,  Biesiada M.,  Pi\'orkowska-Kurpas A.,   Zhu Z.-H.,  2017,
  \mn@doi [Phys. Rev. Lett.] {10.1103/PhysRevLett.118.091102}, 118, 091102

\bibitem[\protect\citeauthoryear{Farrow, Zhu  \& Thrane}{Farrow
  et~al.}{2019}]{Farrow_2019}
Farrow N.,  Zhu X.-J.,   Thrane E.,  2019, \mn@doi [The Astrophysical Journal]
  {10.3847/1538-4357/ab12e3}, 876, 18

\bibitem[\protect\citeauthoryear{Favata}{Favata}{2014}]{Favata_2014}
Favata M.,  2014, \mn@doi [Physical Review Letters]
  {10.1103/physrevlett.112.101101}, 112

\bibitem[\protect\citeauthoryear{Flanagan \& Hinderer}{Flanagan \&
  Hinderer}{2008}]{Flanagan_2008}
Flanagan E.~E.,  Hinderer T.,  2008, \mn@doi [Physical Review D]
  {10.1103/physrevd.77.021502}, 77

\bibitem[\protect\citeauthoryear{Gendreau, Arzoumanian  \& Okajima}{Gendreau
  et~al.}{2012}]{10.1117/12.926396}
Gendreau K.~C.,  Arzoumanian Z.,   Okajima T.,  2012, in Takahashi T.,  Murray
  S.~S.,   den Herder J.-W.~A.,  eds,  Vol. 8443, Space Telescopes and
  Instrumentation 2012: Ultraviolet to Gamma Ray. SPIE, pp 322 -- 329,
  \mn@doi{10.1117/12.926396}, \url {https://doi.org/10.1117/12.926396}

\bibitem[\protect\citeauthoryear{Hallinan et~al.,}{Hallinan
  et~al.}{2017}]{Hallinan_2017}
Hallinan G.,  et~al., 2017, \mn@doi [Science] {10.1126/science.aap9855}, 358,
  1579–1583

\bibitem[\protect\citeauthoryear{Hannuksela, Haris, Ng, Kumar, Mehta, Keitel,
  Li  \& Ajith}{Hannuksela et~al.}{2019}]{Hannuksela_2019}
Hannuksela O.~A.,  Haris K.,  Ng K. K.~Y.,  Kumar S.,  Mehta A.~K.,  Keitel D.,
   Li T. G.~F.,   Ajith P.,  2019, \mn@doi [The Astrophysical Journal]
  {10.3847/2041-8213/ab0c0f}, 874, L2

\bibitem[\protect\citeauthoryear{Haris, Mehta, Kumar, Venumadhav  \&
  Ajith}{Haris et~al.}{2018}]{Haris:2018vmn}
Haris K.,  Mehta A.~K.,  Kumar S.,  Venumadhav T.,   Ajith P.,  2018

\bibitem[\protect\citeauthoryear{Harry \& Hinderer}{Harry \&
  Hinderer}{2018}]{Harry:2018hke}
Harry I.,  Hinderer T.,  2018, \mn@doi [Class.Quant.Grav.]
  {10.1088/1361-6382/aac7e3}, 35, 145010

\bibitem[\protect\citeauthoryear{Hinderer, Lackey, Lang  \& Read}{Hinderer
  et~al.}{2010a}]{PhysRevD.81.123016}
Hinderer T.,  Lackey B.~D.,  Lang R.~N.,   Read J.~S.,  2010a, \mn@doi [Phys.
  Rev. D] {10.1103/PhysRevD.81.123016}, 81, 123016

\bibitem[\protect\citeauthoryear{Hinderer, Lackey, Lang  \& Read}{Hinderer
  et~al.}{2010b}]{Hinderer_2010}
Hinderer T.,  Lackey B.~D.,  Lang R.~N.,   Read J.~S.,  2010b, \mn@doi
  [Physical Review D] {10.1103/physrevd.81.123016}, 81

\bibitem[\protect\citeauthoryear{Kasliwal et~al.,}{Kasliwal
  et~al.}{2017}]{Kasliwal_2017}
Kasliwal M.~M.,  et~al., 2017, \mn@doi [Science] {10.1126/science.aap9455},
  358, 1559–1565

\bibitem[\protect\citeauthoryear{{LIGO Scientific Collaboration}}{{LIGO
  Scientific Collaboration}}{2018}]{lalsuite}
{LIGO Scientific Collaboration} 2018, {LIGO} {A}lgorithm {L}ibrary -
  {LALS}uite, free software (GPL), \mn@doi{10.7935/GT1W-FZ16}

\bibitem[\protect\citeauthoryear{{LIGO Scientific Collaboration} \& {Virgo
  Collaboration}}{{LIGO Scientific Collaboration} \& {Virgo
  Collaboration}}{2019a}]{GCN}
{LIGO Scientific Collaboration} {Virgo Collaboration} 2019a,
  \url{https://gcn.gsfc.nasa.gov/lvc_events.html}

\bibitem[\protect\citeauthoryear{{LIGO Scientific Collaboration} \& {Virgo
  Collaboration}}{{LIGO Scientific Collaboration} \& {Virgo
  Collaboration}}{2019b}]{graceDB}
{LIGO Scientific Collaboration} {Virgo Collaboration} 2019b,
  \url{https://gracedb.ligo.org/latest/}

\bibitem[\protect\citeauthoryear{{LIGO Scientific Collaboration} \& {Virgo
  Collaboration}}{{LIGO Scientific Collaboration} \& {Virgo
  Collaboration}}{2020}]{GW190425_open_data}
{LIGO Scientific Collaboration} {Virgo Collaboration} 2020,
  \url{https://dcc.ligo.org/LIGO-P2000026/public}

\bibitem[\protect\citeauthoryear{Laarakkers \& Poisson}{Laarakkers \&
  Poisson}{1999}]{Laarakkers:1997hb}
Laarakkers W.~G.,  Poisson E.,  1999, \mn@doi [Astrophys.J.] {10.1086/306732},
  512, 282

\bibitem[\protect\citeauthoryear{Lai, Hannuksela, Herrera-Mart\'{\i}n, Diego,
  Broadhurst  \& Li}{Lai et~al.}{2018}]{PhysRevD.98.083005}
Lai K.-H.,  Hannuksela O.~A.,  Herrera-Mart\'{\i}n A.,  Diego J.~M.,
  Broadhurst T.,   Li T. G.~F.,  2018, \mn@doi [Phys. Rev. D]
  {10.1103/PhysRevD.98.083005}, 98, 083005

\bibitem[\protect\citeauthoryear{Li, Mao, Zhao  \& Lu}{Li
  et~al.}{2018}]{Li_2018}
Li S.-S.,  Mao S.,  Zhao Y.,   Lu Y.,  2018, \mn@doi [Monthly Notices of the
  Royal Astronomical Society] {10.1093/mnras/sty411}, 476, 2220–2229

\bibitem[\protect\citeauthoryear{{Li}, {Lo}, {Sachdev}, {Chan}, {Lin}, {Li}  \&
  {Weinstein}}{{Li} et~al.}{2019}]{2019arXiv190406020L}
{Li} A. K.~Y.,  {Lo} R. K.~L.,  {Sachdev} S.,  {Chan} C.~L.,  {Lin} E.~T.,
  {Li} T. G.~F.,   {Weinstein} A.~J.,  2019, arXiv e-prints, \href
  {https://ui.adsabs.harvard.edu/abs/2019arXiv190406020L} {p. arXiv:1904.06020}

\bibitem[\protect\citeauthoryear{Liao, Fan, Ding, Biesiada  \& Zhu}{Liao
  et~al.}{2017}]{Liao_2017}
Liao K.,  Fan X.-L.,  Ding X.,  Biesiada M.,   Zhu Z.-H.,  2017, \mn@doi
  [Nature Communications] {10.1038/s41467-017-01152-9}, 8

\bibitem[\protect\citeauthoryear{Marchant, Breivik, Berry, Mandel  \&
  Larson}{Marchant et~al.}{2020}]{PhysRevD.101.024039}
Marchant P.,  Breivik K.,  Berry C. P.~L.,  Mandel I.,   Larson S.~L.,  2020,
  \mn@doi [Phys. Rev. D] {10.1103/PhysRevD.101.024039}, 101, 024039

\bibitem[\protect\citeauthoryear{Margalit \& Metzger}{Margalit \&
  Metzger}{2017}]{Margalit:2017dij}
Margalit B.,  Metzger B.~D.,  2017, \mn@doi [Astrophys. J.]
  {10.3847/2041-8213/aa991c}, 850, L19

\bibitem[\protect\citeauthoryear{{McIsaac}, {Keitel}, {Collett}, {Harry},
  {Mozzon}, {Edy}  \& {Bacon}}{{McIsaac} et~al.}{2019}]{2019arXiv191205389M}
{McIsaac} C.,  {Keitel} D.,  {Collett} T.,  {Harry} I.,  {Mozzon} S.,  {Edy}
  O.,   {Bacon} D.,  2019, arXiv e-prints, \href
  {https://ui.adsabs.harvard.edu/abs/2019arXiv191205389M} {p. arXiv:1912.05389}

\bibitem[\protect\citeauthoryear{Messenger \& Read}{Messenger \&
  Read}{2012}]{PhysRevLett.108.091101}
Messenger C.,  Read J.,  2012, \mn@doi [Phys. Rev. Lett.]
  {10.1103/PhysRevLett.108.091101}, 108, 091101

\bibitem[\protect\citeauthoryear{Most, Weih, Rezzolla  \&
  Schaffner-Bielich}{Most et~al.}{2018}]{Most:2018hfd}
Most E.~R.,  Weih L.~R.,  Rezzolla L.,   Schaffner-Bielich J.,  2018, \mn@doi
  [Phys. Rev. Lett.] {10.1103/PhysRevLett.120.261103}, 120, 261103

\bibitem[\protect\citeauthoryear{{Mukherjee}, {Wandelt}  \& {Silk}}{{Mukherjee}
  et~al.}{2019}]{2019arXiv190808950M}
{Mukherjee} S.,  {Wandelt} B.~D.,   {Silk} J.,  2019, arXiv e-prints, \href
  {https://ui.adsabs.harvard.edu/abs/2019arXiv190808950M} {p. arXiv:1908.08950}

\bibitem[\protect\citeauthoryear{Murguia-Berthier et~al.,}{Murguia-Berthier
  et~al.}{2017}]{Murguia_Berthier_2017}
Murguia-Berthier A.,  et~al., 2017, \mn@doi [The Astrophysical Journal]
  {10.3847/2041-8213/aa91b3}, 848, L34

\bibitem[\protect\citeauthoryear{Nakamura}{Nakamura}{1998}]{nakamura1998gravitational}
Nakamura T.~T.,  1998, Physical review letters, 80, 1138

\bibitem[\protect\citeauthoryear{Ng, Wong, Broadhurst  \& Li}{Ng
  et~al.}{2018}]{PhysRevD.97.023012}
Ng K. K.~Y.,  Wong K. W.~K.,  Broadhurst T.,   Li T. G.~F.,  2018, \mn@doi
  [Phys. Rev. D] {10.1103/PhysRevD.97.023012}, 97, 023012

\bibitem[\protect\citeauthoryear{Nicholl et~al.}{Nicholl
  et~al.}{2017}]{Nicholl:2017ahq}
Nicholl M.,  et~al., 2017, \mn@doi [Astrophys. J.] {10.3847/2041-8213/aa9029},
  848, L18

\bibitem[\protect\citeauthoryear{{Oguri}}{{Oguri}}{2018}]{2018MNRAS.480.3842O}
{Oguri} M.,  2018, \mn@doi [\mnras] {10.1093/mnras/sty2145}, \href
  {https://ui.adsabs.harvard.edu/abs/2018MNRAS.480.3842O} {480, 3842}

\bibitem[\protect\citeauthoryear{{Oguri}}{{Oguri}}{2019}]{2019RPPh...82l6901O}
{Oguri} M.,  2019, \mn@doi [Reports on Progress in Physics]
  {10.1088/1361-6633/ab4fc5}, \href
  {https://ui.adsabs.harvard.edu/abs/2019RPPh...82l6901O} {82, 126901}

\bibitem[\protect\citeauthoryear{Ohanian}{Ohanian}{1974}]{ohanian1974focusing}
Ohanian H.~C.,  1974, International Journal of Theoretical Physics, 9, 425

\bibitem[\protect\citeauthoryear{Poisson}{Poisson}{1998}]{Poisson:1997ha}
Poisson E.,  1998, \mn@doi [Phys.Rev.D] {10.1103/PhysRevD.57.5287}, 57, 5287

\bibitem[\protect\citeauthoryear{Radice \& Dai}{Radice \&
  Dai}{2019}]{Radice:2018ozg}
Radice D.,  Dai L.,  2019, \mn@doi [Eur. Phys. J.]
  {10.1140/epja/i2019-12716-4}, A55, 50

\bibitem[\protect\citeauthoryear{Radice, Perego, Zappa  \& Bernuzzi}{Radice
  et~al.}{2018}]{Radice:2017lry}
Radice D.,  Perego A.,  Zappa F.,   Bernuzzi S.,  2018, \mn@doi [Astrophys. J.]
  {10.3847/2041-8213/aaa402}, 852, L29

\bibitem[\protect\citeauthoryear{Rezzolla \& Takami}{Rezzolla \&
  Takami}{2016}]{Rezzolla:2016nxn}
Rezzolla L.,  Takami K.,  2016, \mn@doi [Phys.Rev.D]
  {10.1103/PhysRevD.93.124051}, 93, 124051

\bibitem[\protect\citeauthoryear{{Safarzadeh}, {Ramirez-Ruiz}  \&
  {Berger}}{{Safarzadeh} et~al.}{2020}]{2020arXiv200104502S}
{Safarzadeh} M.,  {Ramirez-Ruiz} E.,   {Berger} E.,  2020, arXiv e-prints,
  \href {https://ui.adsabs.harvard.edu/abs/2020arXiv200104502S} {p.
  arXiv:2001.04502}

\bibitem[\protect\citeauthoryear{Samajdar \& Dietrich}{Samajdar \&
  Dietrich}{2019}]{Samajdar:2019ulq}
Samajdar A.,  Dietrich T.,  2019, \mn@doi [Phys. Rev.]
  {10.1103/PhysRevD.100.024046}, D100, 024046

\bibitem[\protect\citeauthoryear{{Sereno}, {Jetzer}, {Sesana}  \&
  {Volonteri}}{{Sereno} et~al.}{2011}]{2011MNRAS.415.2773S}
{Sereno} M.,  {Jetzer} P.,  {Sesana} A.,   {Volonteri} M.,  2011, \mn@doi
  [\mnras] {10.1111/j.1365-2966.2011.18895.x}, \href
  {http://adsabs.harvard.edu/abs/2011MNRAS.415.2773S} {415, 2773}

\bibitem[\protect\citeauthoryear{Smartt et~al.,}{Smartt
  et~al.}{2017}]{Smartt_2017}
Smartt S.~J.,  et~al., 2017, \mn@doi [Nature] {10.1038/nature24303}, 551,
  75–79

\bibitem[\protect\citeauthoryear{Smith et~al.,}{Smith
  et~al.}{2018a}]{smith2018strong}
Smith G.~P.,  et~al., 2018a, arXiv preprint arXiv:1803.07851

\bibitem[\protect\citeauthoryear{Smith, Jauzac, Veitch, Farr, Massey  \&
  Richard}{Smith et~al.}{2018b}]{Smith:2017mqu}
Smith G.~P.,  Jauzac M.,  Veitch J.,  Farr W.~M.,  Massey R.,   Richard J.,
  2018b, \mn@doi [Mon. Not. Roy. Astron. Soc.] {10.1093/mnras/sty031}, 475,
  3823

\bibitem[\protect\citeauthoryear{Smith, Robertson, Bianconi  \& Jauzac}{Smith
  et~al.}{2019a}]{smith2019discovery}
Smith G.~P.,  Robertson A.,  Bianconi M.,   Jauzac M.,  2019a, Discovery of
  Strongly-lensed Gravitational Waves - Implications for the LSST Observing
  Strategy (\mn@eprint {arXiv} {1902.05140})

\bibitem[\protect\citeauthoryear{Smith et~al.,}{Smith
  et~al.}{2019b}]{Smith_2019b}
Smith G.~P.,  et~al., 2019b, \mn@doi [Monthly Notices of the Royal Astronomical
  Society] {10.1093/mnras/stz675}, 485, 5180–5191

\bibitem[\protect\citeauthoryear{Soares-Santos et~al.,}{Soares-Santos
  et~al.}{2017}]{Soares_Santos_2017}
Soares-Santos M.,  et~al., 2017, \mn@doi [The Astrophysical Journal]
  {10.3847/2041-8213/aa9059}, 848, L16

\bibitem[\protect\citeauthoryear{Somiya}{Somiya}{2012}]{Somiya_2012}
Somiya K.,  2012, \mn@doi [Classical and Quantum Gravity]
  {10.1088/0264-9381/29/12/124007}, 29, 124007

\bibitem[\protect\citeauthoryear{Steiner, Hempel  \& Fischer}{Steiner
  et~al.}{2013}]{Steiner_2013}
Steiner A.~W.,  Hempel M.,   Fischer T.,  2013, \mn@doi [The Astrophysical
  Journal] {10.1088/0004-637x/774/1/17}, 774, 17

\bibitem[\protect\citeauthoryear{Takahashi \& Nakamura}{Takahashi \&
  Nakamura}{2003}]{Takahashi_2003}
Takahashi R.,  Nakamura T.,  2003, \mn@doi [The Astrophysical Journal]
  {10.1086/377430}, 595, 1039–1051

\bibitem[\protect\citeauthoryear{Takami, Rezzolla  \& Baiotti}{Takami
  et~al.}{2014}]{Takami:2014zpa}
Takami K.,  Rezzolla L.,   Baiotti L.,  2014, \mn@doi [Phys.Rev.Lett.]
  {10.1103/PhysRevLett.113.091104}, 113, 091104

\bibitem[\protect\citeauthoryear{Tanaka et~al.,}{Tanaka
  et~al.}{2017}]{Tanaka_2017}
Tanaka M.,  et~al., 2017, \mn@doi [Publications of the Astronomical Society of
  Japan] {10.1093/pasj/psx121}, 69

\bibitem[\protect\citeauthoryear{{Tanvir} et~al.,}{{Tanvir}
  et~al.}{2017}]{2017ApJ...848L..27T}
{Tanvir} N.~R.,  et~al., 2017, \mn@doi [\apjl] {10.3847/2041-8213/aa90b6},
  \href {https://ui.adsabs.harvard.edu/abs/2017ApJ...848L..27T} {848, L27}

\bibitem[\protect\citeauthoryear{Thorne}{Thorne}{1983}]{thorne1983theory}
Thorne K.~S.,  1983, in Gravitational radiation. pp 1--57

\bibitem[\protect\citeauthoryear{Troja et~al.,}{Troja
  et~al.}{2017}]{Troja_2017}
Troja E.,  et~al., 2017, \mn@doi [Nature] {10.1038/nature24290}, 551, 71–74

\bibitem[\protect\citeauthoryear{Tsang, Dietrich  \& Van Den~Broeck}{Tsang
  et~al.}{2019}]{Tsang:2019esi}
Tsang K.~W.,  Dietrich T.,   Van Den~Broeck C.,  2019, \mn@doi [Phys.Rev.D]
  {10.1103/PhysRevD.100.044047}, 100, 044047

\bibitem[\protect\citeauthoryear{Veitch et~al.}{Veitch
  et~al.}{2015}]{Veitch:2014wba}
Veitch J.,  et~al., 2015, \mn@doi [Phys. Rev.] {10.1103/PhysRevD.91.042003},
  D91, 042003

\bibitem[\protect\citeauthoryear{Wade, Creighton, Ochsner, Lackey, Farr,
  Littenberg  \& Raymond}{Wade et~al.}{2014}]{Wade_2014}
Wade L.,  Creighton J.~D.,  Ochsner E.,  Lackey B.~D.,  Farr B.~F.,  Littenberg
  T.~B.,   Raymond V.,  2014, \mn@doi [Physical Review D]
  {10.1103/physrevd.89.103012}, 89

\bibitem[\protect\citeauthoryear{Wang, Stebbins  \& Turner}{Wang
  et~al.}{1996}]{PhysRevLett.77.2875}
Wang Y.,  Stebbins A.,   Turner E.~L.,  1996, \mn@doi [Phys. Rev. Lett.]
  {10.1103/PhysRevLett.77.2875}, 77, 2875

\bibitem[\protect\citeauthoryear{{Yu} \& {Wang}}{{Yu} \&
  {Wang}}{2018}]{2018EPJC...78..692Y}
{Yu} H.,  {Wang} F.~Y.,  2018, \mn@doi [European Physical Journal C]
  {10.1140/epjc/s10052-018-6162-9}, \href
  {https://ui.adsabs.harvard.edu/abs/2018EPJC...78..692Y} {78, 692}

\makeatother
\end{thebibliography}

\newpage

\appendix

\section{Gravitational-wave Parameter Estimation}
\label{app:PE}
\noindent The inner product of two real functions $a(t)$ and $b(t)$ are defined as 
\begin{equation}
(a|b)=4 \Re \int_{f_{\textrm{low}}}^{f_{\textrm{high}}} \frac{\tilde a(f)\tilde{b}^{*}(f)}{S_n(f)}df.
\end{equation}
Here, $\tilde a(f)$ is the Fourier transform of $a(t)$, $^*$ denotes complex conjugation and $S_n(f)$ denotes the one-sided power spectral density of the detector noise. $f_{\textrm{low}}$ and $f_{\textrm{high}}$ are the lower-cutoff and higher-cutoff frequency, respectively.

The posterior $p(\vec\theta|d)$ that a signal $h(\vec{\theta})$ with parameters $\vec{\theta}$ is embedded in a given data strain $d$, is given by
\begin{equation}
p(\vec\theta|d,\mathcal{H}) = \frac{\mathcal{L}(d|\vec{\theta},\mathcal{H})p(\vec{\theta}|\mathcal{H})}{p(d|\mathcal{H})}.
\end{equation}

To explore the posterior distribution in the high-dimensional parameter space, we employed the Nested Sampling algorithm as implemented in \textsc{LALInference}~\cite{Veitch:2014wba,lalsuite}.

\section{Methods}\label{app:methods} % Placeholder

In the lensing hypothesis $\HL$, the magnification biases the intrinsic component masses $m_i$ and luminosity distance $D$ to their lensed counterparts. 
Accordingly, the inferred redshift will differ from the true value $z$.

We choose a power-law prior on the magnification and denote the PE-inferred quantities by $m_i^{\rm est}$, $D^{\rm est}$ and $z^{\rm est}$ both in the lensed and unlesed case. Hence, the assumptions which hold under $\HL$ are
\begin{align}\label{eq:lensed_hypothesis_conversion}
\begin{split}
\HL:\thinspace &p(\mu)\propto\mu^{-3},\\
&D = \sqrt{\mu}D^{\rm est}\,,\\
&z = z(\sqrt{\mu}D^{\rm est})\,,\\
&m_i = \frac{m_i^{\rm det}}{1+z} = m_i^{\rm est}\frac{1+z^{\rm est}}{1+z}\,,\\
&\Lambda^{\textrm{est}}_i = \Lambda(m_i) = \Lambda\left(m_i^{\rm est}\frac{1+z^{\rm est}}{1+z}\right)\,,
\end{split}
\end{align}
while under the unlensed hypothesis $\HU$ one finds
\begin{align}
\begin{split}
\HU:\thinspace & m_i = m_i^{\rm est},\\
&D =D^{\rm est}\,,\\
&\Lambda^{\textrm{est}}_i = \Lambda(m_i) = \Lambda(m_i^{\rm est})\,.
\end{split}	
\end{align}

The priors on $m^{\rm est}$ and $D^{\rm est}$ under $\HL$, given $\mu$,  are obtained by change of variables. By means of equations (\ref{eq:lensed_hypothesis_conversion}) one has 
\begin{align}\label{d distrib}
	\begin{split}
	p(D^{\rm est}|\vec{\theta},\mu,\HL) &= p(D^*|\vec{\theta},\mu,\HL)\cdot\left| \frac{\partial D}{\partial D^{\rm est}}\right|\,,\\
	&=p(D^*|\vec{\theta},\mu,\HL)\cdot\mu^{1/2}\,.
	\end{split}
\end{align}
In the above, $\vec{\theta}$ represents all the binary parameters besides masses and distance and $D^*=D(D^{\rm est}, \mu)$, as per the last one of equations (\ref{eq:lensed_hypothesis_conversion}).
The probability that an event at redshift $z=z(D)$ is lensed is measured by the optical depth:
\begin{equation}
\tau(z) = p(\HL|z, \vec{\theta}, \mu)\,.
\end{equation}
The optical depth of lensing is~\cite{Haris:2018vmn}
\begin{equation}
    \tau(z) = 4.17 \times 10^{-6} \left( \frac{D_c(z)}{ {\rm Gpc}}  \right)^3\,
\end{equation}
where $D_c(z)$ is the comoving distance. 
Thus, one has
\begin{align}\label{optical depth prob}
    \begin{split}
    \tau(z) p(z|\vec{\theta,\mu}) &= p(\HL, z|\vec{\theta}, \mu)\,,\\
    &\propto p(z|\vec{\theta}, \mu, \HL)\,.\\
    \end{split}
\end{align}
By means of equation (\ref{optical depth prob}), equation (\ref{d distrib}) becomes:
\begin{align}\label{d distrib optical depth}
    \begin{split}
	p(D^{\rm est}|\vec{\theta},\mu,\HL) &= p(z^{\rm *}|\vec{\theta},\mu,\HL)\cdot\left|\frac{\partial D}{\partial z}\right|\cdot\mu^{1/2}\,,\\
	&\propto \tau(z^*) p(z^{\rm *}|\vec{\theta},\mu,\HL)\cdot \left|\frac{\partial D}{\partial z}\right|\cdot\mu^{1/2}\,,\\
	&\propto \tau(z^{\rm *})p(z^{\rm *}|\HL)\cdot\left|\frac{\partial D}{\partial z}\right|\cdot\mu^{1/2}\,.
	\end{split}
\end{align}
where $z^{\rm *}=z(D^{\rm *})$ and we used the fact that the prior on $z$ is independent of $\vec{\theta}$ and $\mu$.
 In the same fashion, the lensed prior on the masses is:
\begin{align}\label{m distrib}
\begin{split}
&p(m_1^{\rm est}, m_2^{\rm est}|D^{\rm est}, \vec{\theta},\mu,\HL)\\
&= p(m_1^*, m_2^*|D^{\rm est},\vec{\theta},\mu,\HL)\cdot\left| \frac{\partial (m_1,m_2)}{\partial (m_1^{\rm est},m_2^{\rm est})}\right|\\
&=p(m_1^*, m_2^*|D^{\rm est},\vec{\theta},\mu,\HL)\cdot\left(\frac{1+z^{\rm est}}{1+z^*}\right)^2,
\end{split}
\end{align}
where $z^{\rm est}=z(D^{\rm est})$ and $m_i^*=m_i(m_i^{\rm est}, z^{\rm est}, z^*)$.

\subsection{Magnification posterior with mass distributions} 

Here we demonstrate how one can estimate the magnification posterior of a given binary neutron star event, given that it comes from the galactic double neutron star population.
For this purpose, we define the hypothesis $\HLDNS$ to refer to the magnification estimate from the binary \emph{masses}.
I.e., the mass prior $p(m_1, m_2|\HLDNS)$ is the one for galactic double neutron stars, but we make no explicit constraint on the tidal measurements.

In the lensed hypothesis, the joint posterior inferred from a dataset $d$ is:
\begin{align}\label{lensed posteriors}
	\begin{split}
    &p(\mu, D^{\rm est}, m_1^{\rm est}, m_2^{\rm est},\theta|d,\HLDNS)\\
    &\propto\mathcal{L}(D^{\rm est}, m_1^{\rm est}, m_2^{\rm est}, \vec{\theta})p(\mu,D^{\rm est}, m_1^{\rm est}, m_2^{\rm est}, \vec{\theta}|\HLDNS).
	\end{split}
\end{align}
Since the waveform model is unchanged, the likelihood $\mathcal{L}$ is the same under both $\HLDNS$ and $\HU$ and does not depend on $\mu$. The prior is:
\begin{align}\label{lensed priors}
\begin{split}
&p(\mu,D^{\rm est}, m_1^{\rm est}, m_2^{\rm est}, \vec{\theta}|\HLDNS)\\
&=p(m_1^{\rm est}, m_2^{\rm est}|D^{\rm est}, \vec{\theta},\mu,\HLDNS)p(D^{\rm est}|\vec{\theta},\mu,\HLDNS)\\
\qquad&\times p(\vec{\theta}|\HLDNS)p(\mu|\HLDNS),
\end{split}
\end{align}
where we used the fact that $\vec{\theta}$ is independent of $\mu$.

Inserting equation (\ref{lensed priors}) into expression (\ref{lensed posteriors}), we get:
\begin{align}\label{lens_post_1}
	\begin{split}
	    &p(\mu, D^{\rm est}, m_1^{\rm est}, m_2^{\rm est},\vec{\theta}|d,\HLDNS)\\
	&\propto\mathcal{L}(D^{\rm est}, m_1^{\rm est}, m_2^{\rm est}, \vec{\theta})p(m_1^{\rm est}, m_2^{\rm est}|D^{\rm est}, \vec{\theta},\mu,\HLDNS)\\
	\qquad&\times p(D^{\rm est}|\vec{\theta},\mu,\HLDNS)p(\vec{\theta}|\HLDNS)p(\mu|\HLDNS).
	\end{split}
\end{align}
Similarly, the unlensed posterior samples are given by: 
\begin{align}
\begin{split}
	&p(D^{\rm est}, m_1^{\rm est}, m_2^{\rm est},\vec{\theta}|d,\HU)\\
	&\propto\mathcal{L}(D^{\rm est}, m_1^{\rm est}, m_2^{\rm est}, \vec{\theta})p(m_1^{\rm est}, m_2^{\rm est}|D^{\rm est},\vec{\theta},\HU)\\
	\qquad&\times p(D^{\rm est}|\vec{\theta},\HU)p(\vec{\theta}|\HU).
\end{split}
\end{align}
Therefore, one can rewrite equation (\ref{lens_post_1}) as follows:
\begin{align}
	\begin{split}
	&p(\mu, D^{\rm est}, m_1^{\rm est}, m_2^{\rm est},\theta|d,\HLDNS)\\
	&\propto p(D^{\rm est}, m_1^{\rm est}, m_2^{\rm est},\vec{\theta}|d,\HU)p(\mu|\HLDNS)\\
	\qquad&\times \frac{p(m_1^{\rm est}, m_2^{\rm est}|D^{\rm est}, \vec{\theta},\mu,\HLDNS)p(D^{\rm est}|\vec{\theta},\mu,\HLDNS)}{p(m_1^{\rm est}, m_2^{\rm est}|D^{\rm est}, \vec{\theta},\HU)p(D^{\rm est}|\vec{\theta},\HU)}\\
	&= p(D^{\rm est}, m_1^{\rm est}, m_2^{\rm est},\vec{\theta}|d,\HU)\times\mathcal{W},
	\end{split}
\end{align}
where we used the fact that $p(\vec{\theta}|\HLDNS)=p(\vec{\theta}|\HU)$ and the terms in the numerator are computed as prescribed by equations (\ref{d distrib optical depth}) and (\ref{m distrib}).

Since the likelihood is unchanged, the weighting factor amounts to the prior ratio of the two scenarios,
\begin{align}
    \begin{split}
    \mathcal{W} = &\frac{p(m_1^{\rm est}, m_2^{\rm est}| D^{\rm est}, \Vec{\theta}, \mu, \HLDNS)p(\mu|\HLDNS)}{p(m_1^{\rm est}, m_2^{\rm est}|D^{\rm est}, \Vec{\theta}, \HU)p(D^{\rm est}|\Vec{\theta}, \HU)}\cdot\,\\
    &\times \tau(z^{\rm *})p(z^{\rm *}|\HLDNS)\cdot\left|\frac{\partial D}{\partial z}\right|\cdot\mu^{1/2}
    \end{split}
\end{align}
We use a power-law prior on the magnification, $p(\mu|\HLDNS)\propto\mu^{-3}$ in $[2,6000]$. Prior distributions on masses and distance for the lensed case are obtained from the unlensed ones by change of variables from the unlensed to the lensed quantities. The posterior samples and the priors under $\HU$, in turn, are the ones of the \texttt{LALInference} analysis performed by the LIGO and Virgo Collaborations~\cite{PhysRevX.9.031040}. All the other binary parameters are unaffected by the lensing hypothesis and their priors cancel out in the weighting factor. 

\subsection{Magnification posterior with tidal measurements}
To quantify the agreement between the measured tidal deformability and the estimated tidal deformability with a magnification given, we derive the posterior of the magnification $p(\mu|d, \textrm{EOS})$ with a given EOS as follows.

In the following derivation, the other parameters $\vec{\theta}$ are suppressed to ease the notation.
In order to obtain the magnification via tidal measurement, we expand the posterior $p(\mu|d,\textrm{EOS})$ as
\begin{equation}
\begin{aligned}
p(\mu | d, \rm EOS, \HLT) &= p(\mu|\HLT)\int d\tilde{\Lambda}dm^{\rm det}_idD^{\rm est} \\
&\times\mathcal{L}(d|\tilde{\Lambda}, m^{\rm det}_i, D^{\rm est})\\
&\times p(\tilde{\Lambda}, m^{\rm det}_i, D^{\rm est} | \mu, \mathcal{H}_{L}, \textrm {EOS})\\
&= p(\mu|\HLT)\int d\tilde{\Lambda}dm^{\rm det}_idD^{\rm est}\\ 
&\times \mathcal{L}(d|\tilde{\Lambda}, m^{\rm det}_i, D^{\rm est})\\
&\times p(\tilde{\Lambda} | m^{\rm det}_i, D^{\rm est}, \mu, \mathcal{H}_{L}, \textrm{EOS})\\
&\times p(m^{\rm det}_i, D^{\rm est} |  \mu, \mathcal{H}_{L}, \textrm{EOS}).
\end{aligned}
\end{equation}

We notice that the tidal deformability is completely determined with a EOS and source-frame masses (therefore with detector-frame masses and luminosity distance given). Therefore $p(\tilde{\Lambda} | m^{\rm det}_i, D^{\rm est}, \mu, \mathcal{H}_{L}, \textrm{EOS}) = \delta(\tilde{\Lambda} -\tilde{\Lambda}^{\rm est})$, where $\tilde{\Lambda}^{\rm est}$ is the estimated tidal deformability. Therefore, 
\begin{equation}
\begin{aligned}
p(\mu | d, \rm EOS, \HLT) &= \int d\tilde{\Lambda}dm^{\rm det}_idD^{\rm est} \mathcal{L}(d|\tilde{\Lambda}, m^{\rm det}_i, D^{\rm est})\\
&\times \delta(\tilde{\Lambda} -\tilde{\Lambda}^{\rm est}) p(m^{\rm det}_i, D^{\rm est} |  \mu, \mathcal{H}_{L},\textrm{EOS})\\
&\times p(\mu|\HLT)\\
&= \int dm^{\rm det}_idD^{\rm est} \mathcal{L}(d|\tilde{\Lambda}^{\textrm{est}}, m^{\rm det}_i, D^{\rm est})\\
&\times p(m^{\rm det}_i, D^{\rm est} |  \mu, \mathcal{H}_{L}, \textrm{EOS})p(\mu|\HLT).
\end{aligned}
\end{equation}
As the likelihood is unchanged if we switch from $\HLT$ and $\HU$, we then express the likelihoods in terms of the posteriors under $\HU$,
\begin{equation}
\begin{aligned}
p(\mu | d, \rm EOS, \HLT) &\propto \int dm^{\rm det}_idD^{\rm est}\\
&\times\frac{p(\tilde{\Lambda}^{\textrm{est}}, m^{\rm det}_i, D^{\rm est}|d,\HU)}{p(\tilde{\Lambda}^{\textrm{est}}, m^{\rm det}_i, D^{\rm est}|\HU)}\\
&\times p(m^{\rm det}_i, D^{\rm est} |  \mu, \mathcal{H}_{L}, \textrm{EOS})p(\mu|\HLT)\\
&= \int dm^{\rm det}_idD^{\rm est}\\
&\times\frac{p(\tilde{\Lambda}^{\textrm{est}}, m^{\rm det}_i, D^{\rm est}|d,\HU)}{p(\tilde{\Lambda}^{\textrm{est}} | m^{\rm det}_i, D^{\rm est}, \HU)}\\
&\times\mathcal{W}_{\rm EOS}\\
\end{aligned}
\end{equation}
In our study, we sample over the detector-frame masses and individual tidal deformability independently. Based on Eq.~\ref{eq:lambdaTilde}, the prior $p(\tilde{\Lambda}^{\textrm{est}} | m^{\rm det}_i, D^{\rm est}, \HU)$ is the same as the prior $p(\tilde{\Lambda}^{\textrm{est}} | q, \HU)$, where $q \equiv m^{\rm det}_2/m^{\rm det}_1$.
\begin{equation}
\begin{aligned}
p(\mu | d, \rm EOS, \HLT) &\propto \int dm^{\rm det}_idD^{\rm est}\\
&\times\frac{p(\tilde{\Lambda}^{\textrm{est}}, m^{\rm det}_i, D^{\rm est}|d,\HU)}{p(\tilde{\Lambda}^{\textrm{est}} | q, \HU)}\mathcal{W}_{\rm EOS}\\
&= \int dm^{\rm det}_idD^{\rm est} p(m^{\rm det}_i, D^{\rm est} | d, \HU) \\
&\times \frac{p(\tilde{\Lambda}^{\textrm{est}} |  m^{\rm det}_i, D^{\rm est}, d,\HU)}{p(\tilde{\Lambda}^{\textrm{est}} | q, \HU)}\mathcal{W}_{\rm EOS}\\
\end{aligned}
\end{equation}
And finally, we approximate the integral by an average over posterior samples. As a result,
\begin{equation}
\begin{aligned}
p(\mu | d, \textrm{EOS}, \HLT) \propto \left\langle \frac{p(\tilde{\Lambda}^{\textrm{est}}  | d,\HU)}{p(\tilde{\Lambda}^{\textrm{est}} | q, \HU)}\mathcal{W}_{\rm EOS} \right\rangle,
\end{aligned}
\end{equation}
where $p(\tilde{\Lambda}^{\textrm{phase}}|d,\HU)$ is the posterior distribution of the measured tidal deformability, $\tilde{\Lambda}^{\textrm{est}}$ is the estimated tidal deformability with a magnification and a EOS given. $m^{\rm det}_i$ and $D^{\rm est}$ are the observed component masses and the observed luminosity distance, respectively. The weight $\mathcal{W}_{\rm EOS}$ is given by
\begin{equation}
\begin{aligned}
    \mathcal{W}_{\rm EOS} &= \frac{p(m_1^{\rm det}, m_2^{\rm det}| D^{\rm est}, \mu, \HLT, \textrm{EOS} )p(D^{\rm est}|\mu, \HLT)}{p(m_1^{\rm det}, m_2^{\rm det}|D^{\rm est}, \HU)p(D^{\rm est}| \HU)}\\
    &\times p(\mu|\HLT)\\
    &= \frac{p(m_1^{\rm est}, m_2^{\rm est}| D^{\rm est}, \mu, \HLT, \textrm{EOS} )p(D^{\rm est}|\mu, \HLT)}{p(m_1^{\rm est}, m_2^{\rm est}|D^{\rm est}, \HU)p(D^{\rm est}| \HU)}\\
    &\times p(\mu|\HLT).
\end{aligned}
\end{equation}

The difference between $\mathcal{W}_{\rm EOS}$ and $\mathcal{W}$ are the prior on $m^{\rm est}_{1,2}$. For $\mathcal{W}_{\rm EOS}$, the prior on $m^{\rm est}_{1,2}$ is estimated based on a flat prior on the true source component mass to be uniform between $0.5M_{\odot}$ and the maximum mass allowed with a given EOS. While the Galactic double neutron star population is used for the calculation of $\mathcal{W}$ in this paper.

\section{Results with magnification $\mu$ of 100}\label{app:mu100}
\begin{figure*}
    \centering
    \includegraphics[width=\columnwidth]{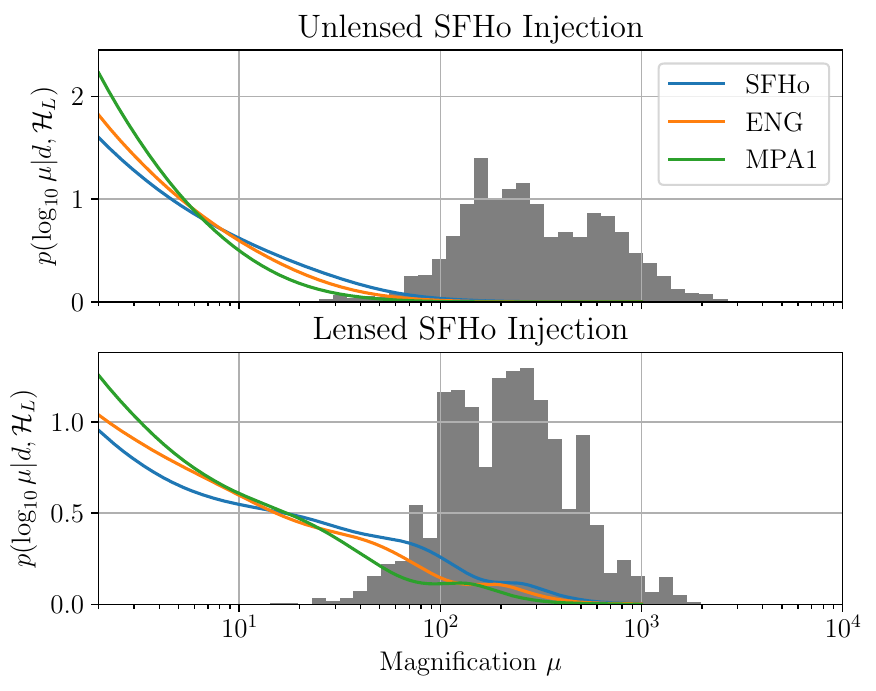} \includegraphics[width=\columnwidth]{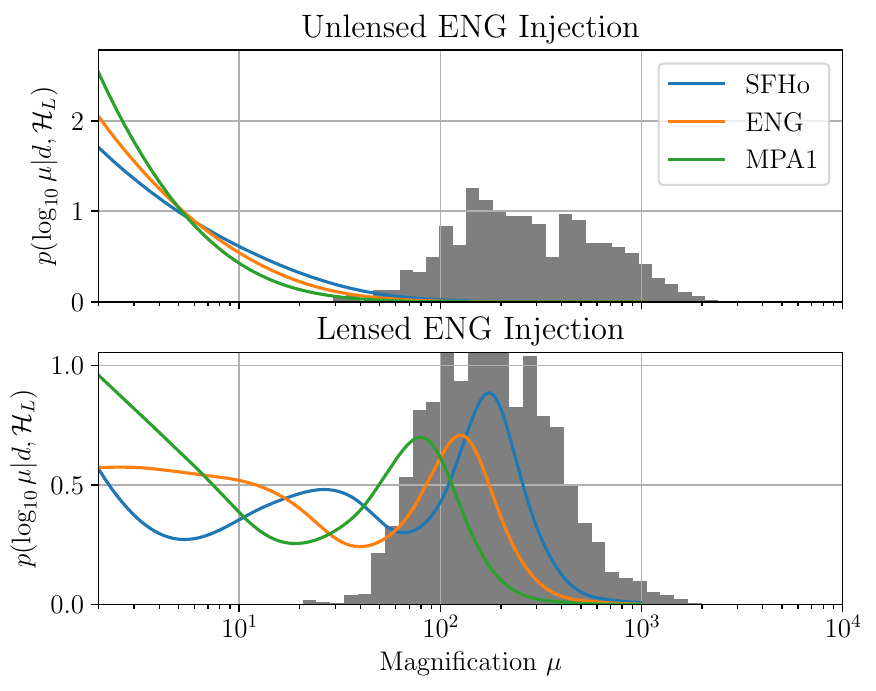}
    \caption{Posterior distribution of magnifications inferred with posteriors of component masses and luminosity distance (gray bins) and that with posteriors of component masses and tidal deformability for given EOSs (colored line) with various injections. 
    We show four different injections: Unlensed SFHo (top left), unlensed ENG (top right), lensed SFHo (bottom left) and lensed ENG (bottom right) injection.
    The posterior of the magnification $\mu$ inferred from the masses and from the tidal deformabilities are giving consistent results for lensed injections. Meanwhile, there exists tension between the posteriors recovered by the two means for unlensed injection. 
    The injected BNS masses are ($m_1=m_2=1.35\, \rm M_\odot$) and ($m_1=m_2=1.58 \, \rm M_\odot$) for the lensed and unlensed binaries, respectively. 
    The binary neutron star is at an observed luminosity distance of $D^{\rm est}=100\,\rm Mpc$, with a signal-to-noise ratio of 25.
    }
    \label{fig:lensingMagnification_mu100}
\end{figure*}

In Fig.~\ref{fig:lensingMagnification_mu100}, we show the magnification posteriors evaluated with the two methods described in Sec~\ref{sec:highmass_binary} with injections tabulated in Tab~\ref{tab:injection_parameters_mu100} given.

We observed that the required magnifications $p(\mu|d,\HLDNS)$, as evaluated from the galactic double neutron star population, are in the $\mu\sim \mathcal{O}(10) - \mathcal{O}(1000)$ range for both the lensed and unlensed injections (Fig.~\ref{fig:lensingMagnification_mu100}, gray bins). 
Meanwhile, the magnifications as estimated from the unbiased tidal deformabilities are different for the two scenarios, favoring the unlensed case for the unlensed injection, and no clear preference for the lensed injection (solid lines, for the SFHo, ENG and MPA1 EOSs).

We find that the two magnification estimates disagree in the unlensed case, ruling out the lensed hypothesis at a log Bayes factor $\log\mathcal{B}^L_U$ of $-0.62(-0.68)$, $-0.77(-0.93)$ and $-1.04(-1.24)$ for SFHo, ENG and MPA1, respectively, for SFHo(ENG) injection. And overlap in the lensed case, showing no clear support on lensed hypothesis at a log Bayes factor $\log\mathcal{B}^L_U$ of $-0.07(0.13)$, $-0.08(0.77)$ and $-0.34(0.01)$ for SFHo, ENG and MPA1, respectively, for SFHo(ENG) injection.

These results show that the lensing hypothesis is disfavoured even for a weaker magnification with a weaker support. Meanwhile, the support for lensed hypothesis under lensed injection is too weak for us to give any statement for it.

\begin{table}
    \centering
    \begin{tabular}{l c c}
    EOS  & Lensed ($\mu=100$) & Unlensed ($\mu=1$)\\
    \hline
    SFHo & (1.35, 432.94) & (1.58, 146.62)\\
    ENG  & (1.35, 644.66) & (1.58, 194.18)\\
    \hline
    \end{tabular}
    \caption{Summary of the source-frame mass and the tidal deformaility of the simulated binary neutron star mergers. Each cell shows the source-frame mass, tidal deformability pair $(m,\Lambda)$ of the injection under different EOS and lensing scenario.}
    \label{tab:injection_parameters_mu100}
\end{table}

%%%%%%%%%%%%%%%%%%%%%%%%%%%%%%%%%%%%%%%%%%%%%%%%%%



 % if your bibtex file is called example.bib

%\begin{thebibliography}{99}
%\end{thebibliography}

%%%%%%%%%%%%%%%%%%%%%%%%%%%%%%%%%%%%%%%%%%%%%%%%%%

% Don't change these lines
\bsp	% typesetting comment
\label{lastpage}
\end{document}